\documentclass[a4paper,fleqn]{cas-sc}

\usepackage{newtxtext}

\usepackage[numbers]{natbib}

\usepackage[autostyle=true]{csquotes}
\usepackage{subfig}
\usepackage{float}
\usepackage{placeins}

\floatstyle{plaintop}
\restylefloat{table}

\usepackage{bookmark}

\makeatletter
\AtBeginDocument{%
  \expandafter\renewcommand\expandafter\subsection\expandafter
    {\expandafter\@fb@secFB\subsection}%
  \newcommand\@fb@secFB{\FloatBarrier
    \gdef\@fb@afterHHook{\@fb@topbarrier \gdef\@fb@afterHHook{}}}%
  \g@addto@macro\@afterheading{\@fb@afterHHook}%
  \gdef\@fb@afterHHook{}%
}
\makeatother

\def\tsc#1{\csdef{#1}{\textsc{\lowercase{#1}}\xspace}}
\tsc{WGM}
\tsc{QE}
\tsc{EP}
\tsc{PMS}
\tsc{BEC}
\tsc{DE}

\let\oldcite\citep
\renewcommand{\citep}[2][]{\mbox{\oldcite[#1]{#2}}}

\usepackage{stackengine}
\usepackage{relsize}
\usepackage{siunitx}
\begin{document}

\captionsetup{labelfont=bf}



\let\WriteBookmarks\relax
\def\floatpagepagefraction{1}
\def\textpagefraction{.001}


\shorttitle{\href{https://doi.org/10.1016/j.oceaneng.2024.119459}{Ocean Engineering 313 (2024) 119459}}

\shortauthors{Draheim et~al.}

\title [mode = title]{Influence of the free surface on turbulent kinetic energy in the wake of a full ship}                      






%
\author[1]{Luise Draheim}[type=editor,
                        auid=000,bioid=1
                        ]

\cormark[1]


\ead{luise.draheim@posteo.de}

\ead[url]{https://www.lemos.uni-rostock.de}

\credit{Methodology, Validation, Formal Analysis, Investigation, Writing - Original Draft, Visualization, Writing - Review and Editing}

\address[1]{University of Rostock, Chair of Modelling and Simulation, Albert-Einstein-Str. 2, 18059 Rostock, Germany}


\author[1]{Nikolai Kornev}[
                        ]

\credit{Supervision}

\cortext[cor1]{Corresponding author}



\let\printorcid\relax

\begin{abstract}
Turbulent kinetic energy (TKE) is a measure for unsteady loads and important regarding the design of e.g. propellers or energy-saving devices. While simulations are often done for a double-body, using a symmetry condition, experiments and the final product have a free surface. Simulations with and without free surface are carried out for the Japan Bulk Carrier, comparing TKE in the vortex cores. The reliability of finding the vortex centers is discussed. As the fine meshes show an unexpected trend for the TKE, a detailed investigation is done, mainly to exclude method-related drawbacks from using a hybrid URANS/ LES model. It is found that a shift in vortex-core positions distorts the results whereby the experimental center positions which are referenced are questionable. Using a fixed position for all cases improves comparability and gives a different picture. Thereupon the medium meshes were enhanced in such a way that one of the refinement boxes was extended further forward, now showing much better agreement with the fine meshes. TKE is then portrayed as integral quantity and shows no significant difference between the simulations with and without free surface. However, the structure itself is influenced by the surface in a way which alters local characteristics.

\end{abstract}







\maketitle


\section{Introduction}

While energy-efficiency became more and more important over the last years \citep{BAL18}, especially regarding the ecological footprint, there is ongoing development of optimizations and devices which could lead to a reduction of energy demand. Due to well-advanced research, improvements are often within a very small range, raising a question if simulated/ tested results will lead to real-scale enhancements. With ongoing complexity of simulations and increased computing effort, simplifications are applied where possible. For slow-steaming ships in calm water, often the free surface is neglected and replaced by a mirror plane. However, also at low speed a stationary wave is formed behind the ship in the wake and alters the surface. Nowadays multiple software packages are capable of including the free surface, either through an interface tracking or interface capturing method \citep[p. 602 ff.]{FER20}. Studies on underwater bodies close to the surface reveal disturbances like a change of the wake flow-field structure and position, or an influence on drag coefficients even at small Reynolds numbers \citep{KIL22} \citep{JAG10} \citep{NEM15}. For large ships like e.g. bulkers, differences can be detected regarding the velocity contours, but for usual cases they do not have an effect on important parameters like resistance and/ or are not significant. Considering more extreme cases, like e.g. waves, it was shown, that it could have a significant influence on the performance of an energy-saving device (ESD), to the point where the thrust loss with ESD was higher than without \citep{BAK20}. Also, the free surface is assumed to be responsible for differences in mean pressure values between measurement and CFD, which equal the hydrostatic part of the pressure due to the sensors being submerged in the waves generated by the hull \citep{ANS23}. Regarding energy efficiency, another important finding is, that detailed flow features like peak loadings on propellers may account for noticeable changes in thrust \citep{ABB15}. A measure for these loads is the turbulent kinetic energy (TKE). Not only thrust can be affected by peak loadings, but also the structural integrity.\\
The main purpose of this paper is to study the effect of the free surface on the magnitude and spatial distribution of turbulent kinetic energy in the wake of the JBC test ship, and to draw conclusions as to how important the consideration of the interface is for cases where the results could be affected by non-stationary loads.\\
Next up the motivation and state of knowledge will be explained in more detail. The underlying theory is then described, followed by the mesh set-up and case specifications. Within the results section, some general verification data is presented, followed by a grid convergence discussion and the first results. Since some doubts arose about the results, further simulations and analyses were conducted, which are presented afterwards. In the discussion, the reasons are addressed in depth and interim results and final findings are explained.

\subsection{Motivation}

Non-stationary loads in the wake of a ship, e.g. on the propeller, are generally caused by three factors:

\begin{enumerate}
	\item unsteadiness due to variable velocity of the liquid relative to the vessel, which in turn can be caused by the variability of the vessel's speed, waves and free stream turbulence
	\item mean non-uniformity of the wake
	\item turbulence of the wake, which is especially strong for full vessels such as tankers and bulk carriers.
\end{enumerate}

While the first point is not considered in this article, the second factor can be studied using the common Reynolds-Averaged Navier-Stokes (RANS) technique. Analysing the third factor, the determination of the turbulence structures is required, which is only possible within the framework of scale-resolving methods, such as Large Eddy Simulation (LES) and hybrid URANS/ LES \cite{BHU12}, \cite{FUR16}, \cite{XIN12}.

As it was shown in the calculations of the Tokyo 2015 workshop (Tokyo’15), the level of turbulent kinetic energy in a number of sections behind the Japan Bulk Carrier (JBC) test case, determined by different (U)RANS techniques, differs from the experiment by an order of magnitude \cite{HIN21}.\\
The unsteady loadings on propellers were studied using (U)RANS and hybrid approaches in our previous works \cite{ABB15}. It has been shown that the fluctuations in the thrust of the propeller, determined using URANS, are regular over time and have a relatively low magnitude. On the contrary, thrust fluctuations, determined both using the hybrid method and in the experiment, are irregular with large peaks and increased magnitude. This highlights the need to apply hybrid methods and resolve turbulence in the wake. As far as we know, studies of turbulence with scale-resolving methods in the wake were conducted only using a single-phase flow model. The effect of the free surface on the TKE in the wake has not yet been investigated.

\subsection{Literature}

While for Tokyo'15 the influence of the free surface was found to be \enquote{not negligible but not large} \citep[p. 184]{HIN21}, a real-scale benchmark study by \textit{Andersson et al.} \citep{AND22} found no significant correlation. However, let us have a closer look at these two findings. First, during Tokyo'15, the JBC was examined. Experiments and simulations were done for a low Froude number, which typically allows for making use of the double-body approach and single-phase simulations. While this is true for parameters like resistance, only few simulations were done to figure out the influence of the free surface on the wake. None of them was done using a scale-resolving method, and the TKE was not examined. The study by \textit{Andersson et al.} had several participants which did CFD simulations of the Kriso Container Ship (KCS), some including free surface. Correlations were searched between the influence of the free surface and thrust at zero wind and waves. While they did not find a relation, they emphasized a lack of detailed validation data and pointed to a possible case-dependency. Additionally, again none of the simulations was done using a scale-resolving approach.\\
There have been early attempts to show the influence of the free surface on the turbulent kinetic energy, mostly in experiments.\\
A very general approach to start with is the elementary difference of the pressure distribution.

\begin{figure}[!htbp]
	\centering
		\includegraphics[scale=.75]{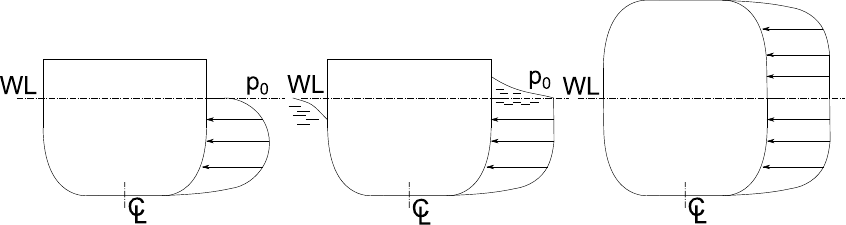}
	\caption{Ship cross section with a) left: undisturbed surface, b) middle: waves, and c) right: double body \citep{TRE76}, where {\raisebox{ .55ex}{C}}\kern -0.5em{L}: centerline, WL: waterline and $p_0$: atmospheric pressure.}
	\label{fig:1}
\end{figure}

\FloatBarrier

Figure~\ref{fig:1} illustrates, why the double-body approach is considered reasonable for low Froude numbers, where the surface is hardly deformed. A moving ship always radiates waves, changing the surface elevation and thus the pressure distribution. Assuming small motions and changes, the implications by the change in density from water to air are similar to a rigid plate at the free surface. Using a symmetry condition where only the normal forces remain, the body is virtually mirrored and the forces are cancelling each other out.\\
The Froude number is considered low if $Fn \leq 0.15$, although for some ships (like e.g. the JBC) with sudden changes in hull form even smaller values are needed. The differences in pressure distribution are small, especially close to the hull where relevant phenomena develop. However, even slow forward speed leads to a deformation of the surface, especially for full ships or vessels with sudden change of the hull form.\\
The expectation then is a change in pressure distribution and additional effects e.g. from a varying velocity profile.\\
\textit{Maheo} did some experiments with a flat plate and an undisturbed free surface and investigated the wake with Particle Image Velocimetry (PIV). He described the origin of secondary flows which emerge because of a mixed boundary conditions due to the free surface \citep[pp. 99-100]{MAH99}. A formula was developed which could show, that \enquote{free-surface interaction serves to redistribute the energy from the surface-normal to tangential velocity fluctuations} \citep[p. 118]{MAH99}.\\
Often turbulence due to free surface is associated with wave-breaking, but \textit{Babanin et al.} could show, that for short and steep laboratory waves it exists solely because of the orbital motion. By creating unforced waves (mechanically, no wind, deep water) they eliminated sources of shear stress and background turbulence. Measurements were done close to the surface, right below the wave troughs. The observed occurrence can affect the subsurface boundary layer, with dissipation rates of the non-breaking turbulence (locally at the rear face of the wave) found to be similar to breaking-in-progress rates at the front face of waves \citep{BAB09}. \\
\textit{Kahraman et al.} could show, that the free surface can have an effect on the velocity field. As they did CFD simulations, the calculations using a Volume-of-Fluid (VOF) method were in better agreement with the experiments. Regarding TKE, a higher amount due to the free surface leads to slowing down the flow, even at low Froude numbers \citep{KAH20}. However, the calculations were done using URANS only, thus the result is still remarkable, but does not allow a reliable insight into detailed flow features.\\
In 2022 the author presented results from a study, which included the very first tentative interim findings regarding the scientific problem discussed in this paper. Due to the preliminary character of that study, only a single mesh size was investigated, corresponding to the (initial) medium mesh herein. The main noteworthy differences are a larger Courant number and shorter runtime in 2022 \citep{DRA22}. Following various recommendations, in the present paper smaller Courant numbers were chosen considering case set-ups for scale-resolving methods (\verb|maxCo|) and simulations with an interface (\verb|maxAlphaCo|), and the averaging time was doubled. In hindsight, the initial choices i.a. led to equivocal identification of more than one possible vortex core. Due to not having investigated different grid sizes at that time, possible flaws of the mesh configuration were unknown. Differences in results and conclusions from 2022 stem from the previously described changes, underlining the importance of grid studies.


\section{Theory}

\subsection{Turbulence model}

For vessels, fully turbulent flow is assumed at $Re \geq 10^6$, whereas usual ships have a design speed which corresponds to ranges of $10^7 \leq Re \leq 10^9$. This is important, as hybrid models work best (or only) if there is a fully developed turbulent flow and strong separations in the boundary layer \citep{TEM05}. As turbulence is a highly unsteady phenomenon, transient CFD codes are necessary to describe the flow.\\
The momentum equations valid for URANS and LES reads in general form:
\begin{equation}
	\label{eq:momEq}
	\frac{\partial \overline{u_i}}{\partial t} + \overline{u_j}\frac{\partial \overline{u_i}}{\partial x_j} + \frac{1}{\rho}\frac{\partial \overline{p}}{\partial x_i} = \frac{\partial}{\partial x_j}\left[\nu\left(\frac{\partial \overline{u_i}}{\partial x_j} + \frac{\partial \overline{u_j}}{\partial x_i}\right) + \tau_{ij}^h \right] + \overline{f_i}
\end{equation}

with the pressure $p$, density $\rho$ and the source term $f_i$. While the overline means filtering for LES, it is time-averaging for URANS. The term $\tau_{ij}^h$ also has different meanings; for LES it is the subgrid stress $\tau_{ij}^h = \tau_{ij}^{SGS} \left(\Delta, \overline{u_i}, C \right)$, for URANS it is the Reynolds stress $\tau_{ij}^h = \tau_{ij}^{RANS} \left(\overline{u_i}, k, \epsilon, C \right)$, with the turbulent kinetic energy $k$, dissipation rate $\epsilon$, characteristic cell size $\Delta$ and $C$ a (set of) model constant.\\
All simulations are based on the k-$\omega$-SST model by \textit{Menter} \citep{MEN93} which is a two-equation model based on the Boussinesq hypothesis. The Reynolds stress is now defined assuming isotropic turbulence:
\begin{equation}
	\label{eq:boussi}
	- \overline{u_{i}' u_{j}'} = \nu_{t} \overline{S_{ij}} - \frac{2}{3} \delta_{ij} k
\end{equation}

with the strain-rate tensor $S_{ij} = \frac{1}{2} \left( \frac{\partial \overline{u_i}}{\partial x_j} + \frac{\partial \overline{u_j}}{\partial x_i}\right)$ and Kronecker delta $\delta_{ij}$. The turbulent viscosity $\nu_{t}$ is here defined with a transfer function $F_{2}$, a constant $a_1 = 0.31$, the specific dissipation rate $\omega$ and the vorticity magnitude $S = \sqrt{2S_{ij}S_{ij}}$:
\begin{equation}
\nu_{t} = \frac{a_{1} k}{max(a_{1} \omega , S F_{2})}
\end{equation}

\subsubsection{SLH model}

The LeMoS-hybrid (LH) model was developed by \textit{Kornev et al.} \citep{KOR11} and takes advantage of the similar formulation for URANS and LES, applying a blending function $f(x,t)$ to $\tau_{ij}^h$. Using the k-$\omega$-SST model and dynamic Smagorinsky model, the stress can be expressed through the viscosity:
\begin{equation}
	\label{eq:stressThroughViscosity}
	\tau_{ij}^h = - 2\nu_h\overline{S_{ij}}
\end{equation}

The hybrid viscosity can be written as sum of turbulent and subgrid-scale viscosity:
\begin{equation}
	\label{eq:hybridViscosity}
	\nu_h = f\nu_t + (1 - f)\nu_{SGS}
\end{equation}

The value of the blending function varies between zero and one, depending on the parameter $h$:
\begin{equation}
	\label{eq:slhCriterion}
	h(x,t) = \frac{L(x,t)}{\Delta(x)}
\end{equation}

which emerges from the integral length scale $L = C_L \frac{\sqrt{k}}{\beta^*\omega}$ (with $\beta^* = 0.09$ and $C_L = 1$) \citep[p. 703]{MOU16} and the characteristic cell size $\Delta = \sqrt{0.5(\Delta_{max}^2 + V^{2/3})}$ with maximum cell edge length $\Delta_{max} = max(\Delta_x,\Delta_y,\Delta_z)$ and cell volume $V$.
To reduce grid-induced separation (GIS) a shielding function was introduced by \textit{Shevchuk et al.} \citep{SHE18}, based on the function $r_d$ by \textit{Gritskevich et al.} \citep{GRI12}, but using $S^2$ and $\Omega^2$ instead of $\sqrt{u_{ij}u_{ij}}$.\\
The code was validated i.a. for the asymmetric plane diffuser flow, the turbulent channel flow, the KVLCC2 benchmark case, and successfully applied to two more benchmark models (DTC and JBC) \citep{KOR11} \citep{ABB15} \citep{ABB16a} \citep{SHE18} \citep{KOR18}.

\subsection{Free-Surface-Capturing Method}

Multiple approaches exist to capture the free surface within CFD, which differ fundamentally in their concept. As hereinafter the OpenFOAM framework is used, two methods are basically available: algebraic VOF and geometric VOF. The latter is implemented through the \verb|interIsoFoam| solver, making use of the \verb|isoAdvector| code \citep{ROE16}. Due to the higher demand of computing capacity, the \verb|interFoam| solver was chosen instead, which is based on a phase-fraction approach. The density $\rho$ from the momentum equation (eq. \ref{eq:momEq}) is here defined as
\begin{equation}
	\label{eq:rhoInterFoam}
	\rho = \alpha\rho_1 + (1 - \alpha)\rho_2
\end{equation}

where $\alpha$ equals 1 in the fluid with the density $\rho_1$, and zero in the fluid $\rho_2$. At the interface $\alpha$ varies between these two value. An additional equation for $\alpha$ has to be solved:
\begin{equation}
	\label{eq:alphaInterFoam}
	\frac{\partial \alpha}{\partial t} + \frac{\partial (\alpha u_j)}{\partial x_j} = 0
\end{equation}

The boundedness of scalar fields like the phase fraction is realized with the semi-implicit multi-dimensional limiter for explicit solution (MULES) \citep{DAM13} \citep{ALM18}.

\subsection{Evaluation}

To analyse the results, the procedures are largely adopted from the Tokyo'15 workshop. A sufficiently large part of the sector around the core is extracted manually. This is done, as the extreme values of the parameters which are needed later for the determination of the center coordinates are located in the separation region and/ or boundary layer. The part is only used to find the core position, all further data is taken from the full dataset. To get started, the vortex cores are determined, finding the maximum axial vorticity $\Omega_x$:
\begin{equation}
	\label{eqn:vorticity}
	\Omega = \nabla \times U
\end{equation}

From the center a horizontal and vertical line is drawn where the TKE is plotted. The turbulent kinetic energy is a measure for the intensity of the turbulence, and can be interpreted as mean kinetic energy per unit mass related to the vortices, respectively the velocity fluctuations \citep[p. 50]{VER07}. In terms of CFD, the TKE from unsteady (hybrid) simulations is composed of a resolved and the modelled part $k_{mean}$:
\begin{equation}
	\label{eqn:TKEtotal}
	k_{total} = \frac{1}{2}\left( \overline{u_{xx}'}^2 + \overline{u_{yy}'}^2 +\overline{u_{zz}'}^2 \right) + k_{mean}
\end{equation}

The share of resolved TKE can be calculated as:
\begin{equation}
	\label{eqn:resTKE}
	k_{res} = \frac{\frac{1}{2}\left( \overline{u_{xx}'}^2 + \overline{u_{yy}'}^2 +\overline{u_{zz}'}^2 \right)}{\frac{1}{2}\left( \overline{u_{xx}'}^2 + \overline{u_{yy}'}^2 +\overline{u_{zz}'}^2 \right) + k_{mean}}
\end{equation}

During runtime output was created, e.g. the scalar and vector fields at the sections, which later was processed with Python. The identification of the vortex core is done visually, checking axial vorticity, mean Q-criterion and $\lambda2$-criterion. The Q-criterion is the second invariant of the velocity gradient tensor and widely used to create impressions of 3D vortical structures with a contour filter.
\begin{equation}
	\label{eqn:qcrit}
	Q = \frac{1}{2} \left( \Omega_{ij}\Omega_{ij} - S_{ij}S_{ij} \right)
\end{equation}

The $\lambda2$-criterion can also be used to identify eddies, but like most vortex-identifying methods a disadvantage is the assumption of only one single main vortex \citep{JIA05}. On the other hand, unlike vorticity, it is said to \enquote{detect} the shear layer thus not erroneously detecting a vortex there. It is derived from the second largest eigenvalue of the sum of the square of the symmetric and antisymmetric parts of the velocity gradient tensor $\left( \Omega_{ij}\Omega_{ij} + S_{ij}S_{ij} \right)$ \citep{JEO95}.


\section{Numerical Set-Up}

\subsection{Model Geometry}

All simulations are using geometry and data of the JBC model (see table \ref{tab:1}) by the National Maritime Research
Institute Japan (NMRI). To achieve utmost comparability between the 1P- and 2P-cases, the hull is fixed at zero sinkage and trim. This should exclude further external influences, and is not considered critical, as the attitude scarcely changes \citep{AND15}. The simulations are done for the bare hull in calm water, without propeller, rudder or ESD. Although turbulence is unsteady and considered to be asymmetrical, the analysed regions are assumed to be not substantially influenced by these effects, and thus only half of the body is modelled, using a symmetry plane.

\begin{table}[H]
\centering
\caption{Main particulars of JBC with model scale 1:40 \citep{NAT15a}}
\label{tab:1}
\begin{tabular}{llll}
\hline
Length between perpendiculars & $L_{PP}$ & [$m$] & 7 \\
Maximum beam at waterline   & $B_{WL}$ & [$m$] & 1.125 \\
Depth   & $D$ & [$m$] & 0.625 \\
Draft   & $T$ & [$m$] & 0.4125 \\
Displacement (Hull) & $\nabla$ & [$m^3$] & 2.787 \\
Wetted Surface Area (Hull) & $S_{0\_w/oESD}$ & [$m^2$] & 12.223 \\
Block coefficient & $c_B$ & [-] & 0.858 \\
Longitudinal center of buoyancy & $L_{CB}$ & [$\%L_{pp}$], fwd+ & 2.5475 \\
Vertical center of gravity & $KG$ & [$m$] & 0.33225 \\
\hline
Velocity & $U$ & [$m/s$] & 1.179 \\
Froude number & $Fn$ & [-] & 0.142 \\
Reynolds number & $Re$ & [-] & $7.46\cdot 10^6$ \\
\hline
\end{tabular}
\end{table}

\FloatBarrier

The coordinate origin of the ship is set to be on keel, at centerline and at $50\%$ $L_{PP}$. The length over all ship (LOA) is $\qty{7.274}{m}$. Other experimental values which are adopted for the simulations are the gravitation constant $g = \qty{9.80}{m/s^2}$, kinematic viscosity $\nu = 1.107 \cdot 10^{-6}\qty{}{m^2/s}$ and water density $\rho = \qty{998.2}{kg/m^3}$.

\FloatBarrier


\subsection{Mesh Set-Up}

Different approaches exist to determine a sufficient grid spacing for LES \citep{LIE17}. The author started from the Kolmogorov length scale $\eta$ which is used to determine cell sizes of DNS (direct numerical simulation) cases. As most of the dissipation happens at larger scales than the Kolmogorov length scale, even for DNS the scale can be set to $5\eta ... 15\eta$, and if the main characteristics are well met by the calculation the factor can even be set to 100 \citep[p. 112]{VER07}. Unfortunately, most of the approximations in CFD are for standard cases like a tube (inner flow) or a flat plate (outer flow), thus there are a lot of discussions regarding how to choose e.g. the inflow turbulence or the reference length for a ship. The author chose to use the ship length as reference length, and as DNS has much higher demands, the factor 100 is considered to be sufficient for LES. Miscellaneous simulations proved this decision true, while e.g. the Taylor microscale \citep[p. 200]{POP00} not yielded the desired results. Please keep in mind that this approach only serves as a starting point.\\
In addition, there are other problems:
\begin{itemize}
	\item restriction of minimum cell size due to size of first layer $\rightarrow$\\
	The advantage of a hybrid method lies i.a. in making use of wall functions. It was found, that using or waiving wall functions has no significant influence on the results \citep[p. 145, pp. 179-180]{HIN21}. While inside the community different opinions exist regarding optimal $y^+$ values in general, it was found during Tokyo’15 for the JBC that $y^+ = 50...60$ worked well, and thus the first-cell size defines the minimum size of the volume grid.
	\item restriction of minimum cell size due to meshing of free surface $\rightarrow$\\
	Doing (hybrid) LES  with OpenFOAM, it is strongly advised to use structured or fully hexahedral meshes due to the algorithms, and especially the filters, having been developed for hexahedral cells \citep{OPE16}. One drawback of hexahedral grids is that refinement only works by halving the cell size immediately, withholding the possibility to smoothly change to the next level. Having a free surface, it is crucial to sufficiently discretize the area around the waterline, especially adequately well ahead of the ship and in zones of steep wave angles. Latter requires in many cases preferably equilateral cells. From the experiments the biggest wave height was gathered and taken as initial value for the smallest (equilateral) cell size in the waterline. Starting from there, the other cell sizes are determined. Thus, the smallest volume-grid cell is bound to be a negative power of the free-surface-cell size.
	\item problems of \verb|interFoam| with values $y^+ < 30$ $\rightarrow$\\
	While unfortunately not explicitly stated anywhere, the author found some threads and talked to colleagues about \verb|interFoam| not working with low values of $y^+$ for some cases. Problems were reported for e.g. the Duisburg Test Case (DTC), but it is not yet thoroughly investigated which conditions (e.g. solver) cause or influence the result. If the (mean) value of $y^+$ falls below the limit, random pressure peaks emerge randomly and let the simulation crash within a short time or give completely odd results. The reason seems to be an unstabilised solution, perhaps as discussed by \textit{Henrik Rusche} \citep[pp. 146-147]{RUS03}. A possible fix is to use small values for $y^+$ only further away from the waterline. The author tested this option having $y^+ \approx 1$ on the strut (and only there), yielding acceptable results. Still, the RANS was quite unstable, and the transient simulations needed a very low Courant number.
\end{itemize}

\begin{wrapfigure}{l}{0pt}
	\includegraphics[width=.5\linewidth]{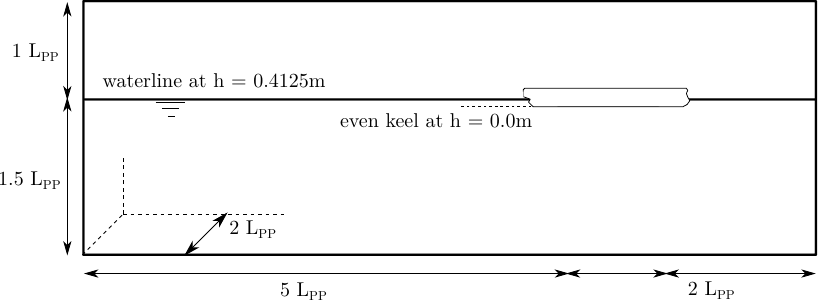}
	\caption{Domain size.}
	\label{fig:2}
	\vspace{-1em}
\end{wrapfigure}

\begin{figure}[!htp]
	\centering
	\subfloat[Discretization of waterline.]
	{\includegraphics[width=.5\linewidth]{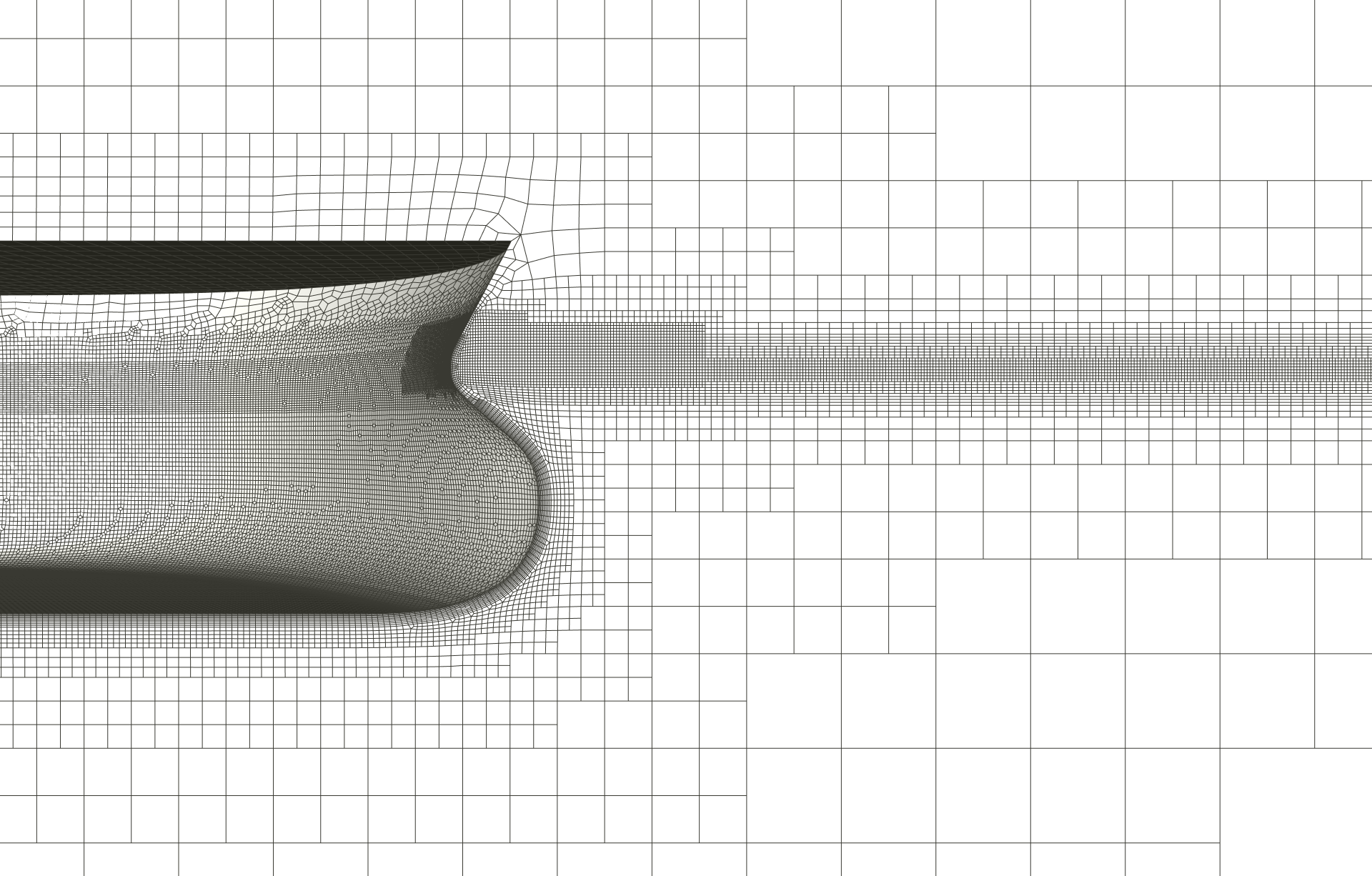}
	\label{fig:3a}}
	\subfloat[Stern refinements.]
	{\includegraphics[width=.5\linewidth]{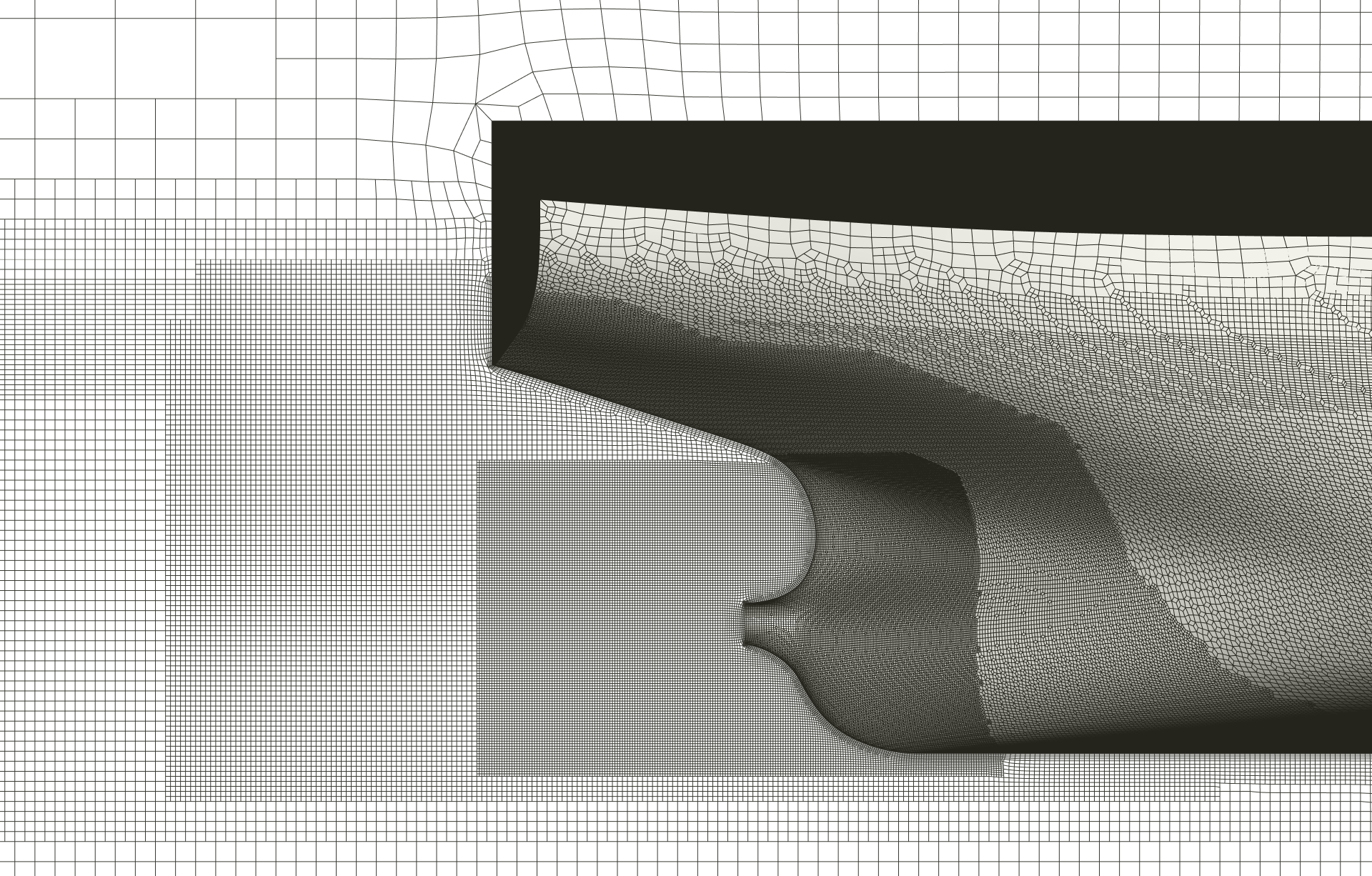}
    \label{fig:3b}}\\
    \subfloat[Top view at waterline on aft ship.]
	{\includegraphics[width=.33\linewidth]{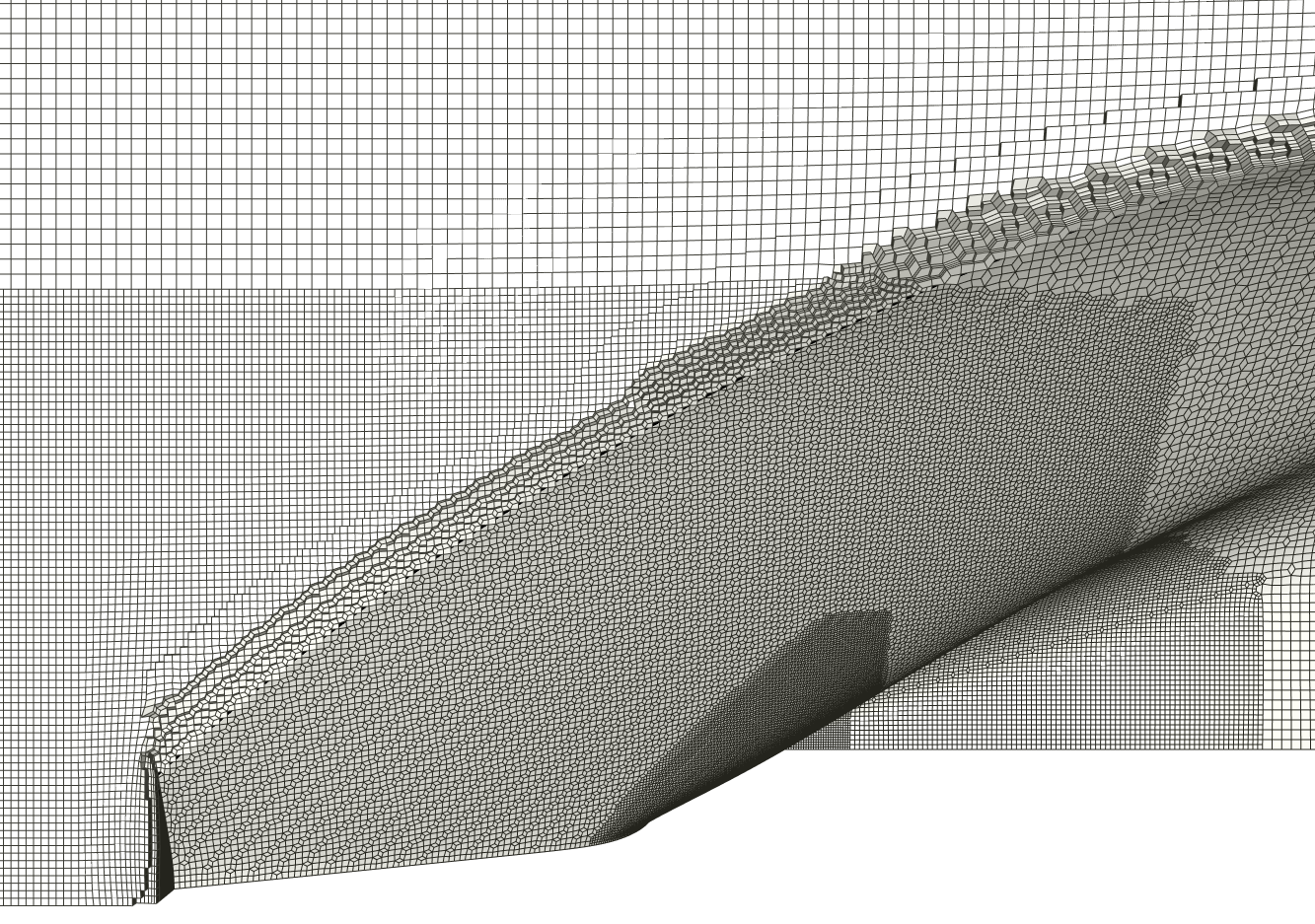}
    \label{fig:3e}}
    \subfloat[Top view at waterline on fore ship.]
	{\includegraphics[width=.33\linewidth]{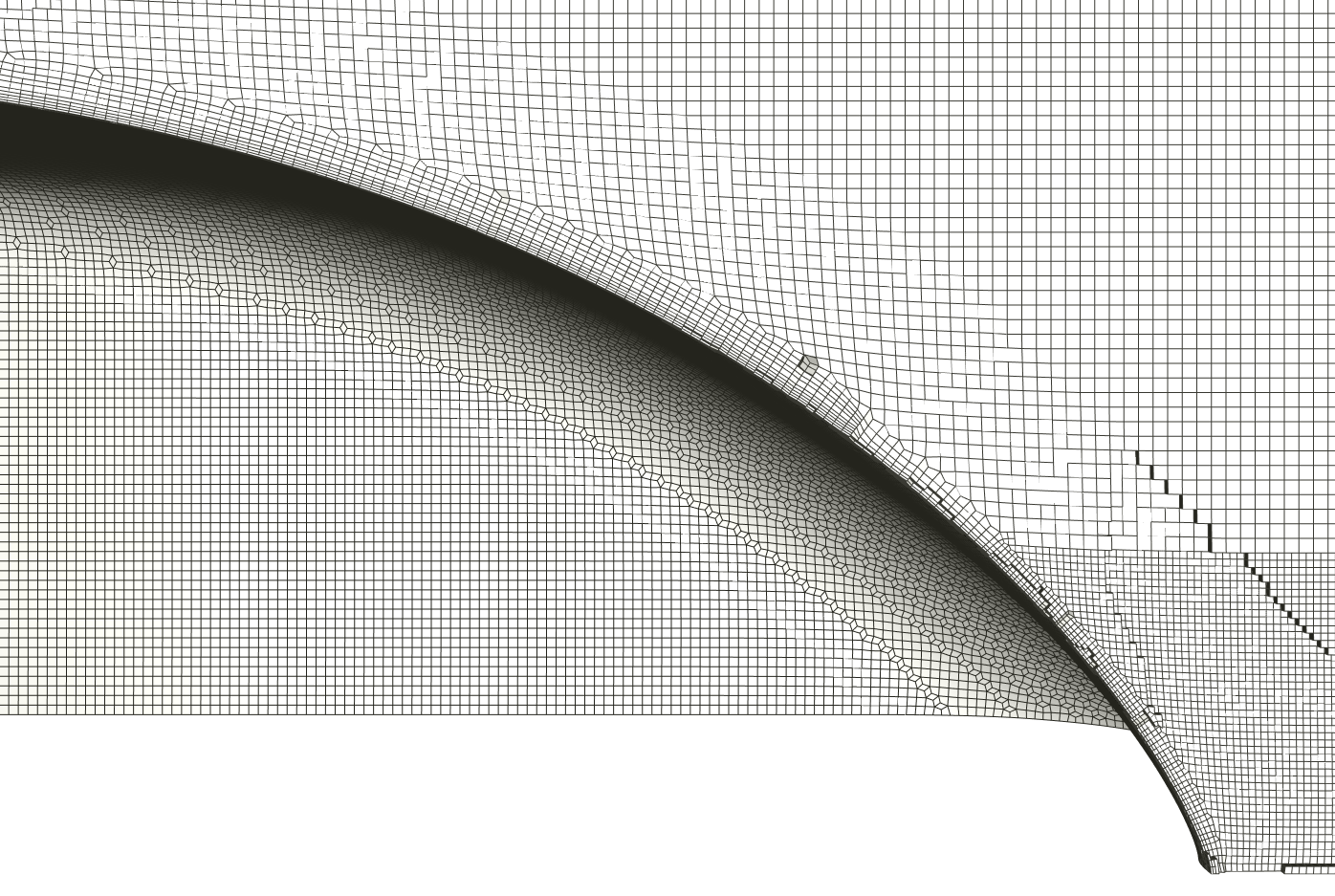}
    \label{fig:3d}}
	\subfloat[Top view at waterline.]
	{\includegraphics[width=.33\linewidth]{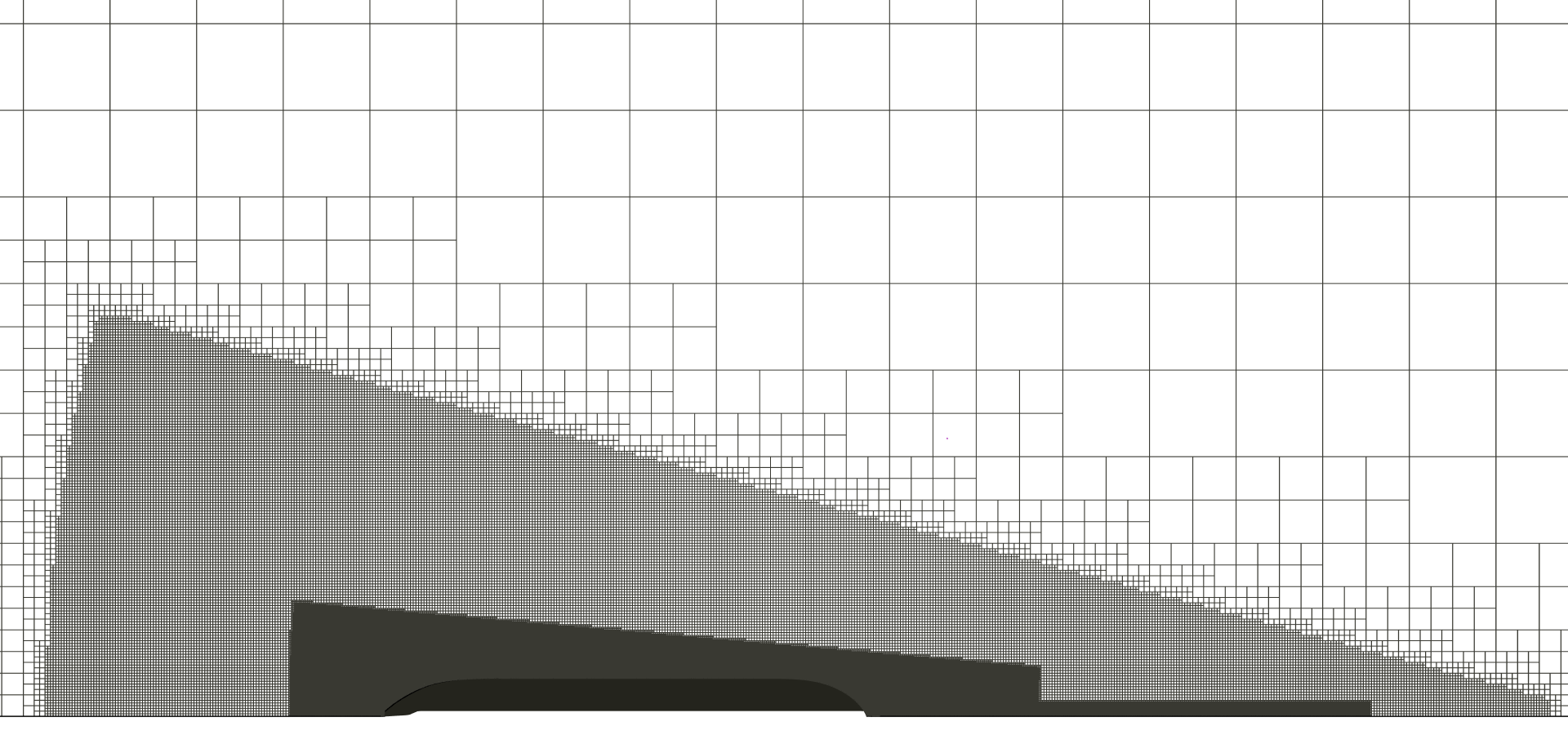}
	\label{fig:3c}}
    \caption{Medium 2P mesh of JBC.}
    \label{fig:3}
\end{figure}

Following from these points, the main restriction is $y^+$, followed by the cell size of the free surface. Different meshes are created for single-phase (1P) and two-phase (2P) simulations. The starting configurations include a medium and a fine mesh in each case. The 2P domain is shown in figure~\ref{fig:2}, details in figure~\ref{fig:3} and the boundary conditions in table~\ref{tab:2}. The only difference to the 1P mesh is the cut at water level and omitting the refinement of the waterline (Kelvin pattern and far ahead of the ship). A first rough RANS simulation, as well as the experimental results, show the largest wave height to be around $\qty{0.1}{m}$. Supplementary test simulations gave the best results in terms of agreement with the experimental values, e.g. on wave elevation, when the waterline was subdivided into cells of $\qty{0.005}{m}$ height. On that basis, the initial cell size is $\qty{1.28}{m}$, and nine (medium mesh), respectively ten (fine mesh) refinement steps are used. The approximated smallest necessary cell size amounts to $\qty{0.0025}{m}$ (medium mesh) and $\qty{0.00125}{m}$ (fine mesh). The length scale based on DNS grid size amounts to $\qty{0.0049}{m}$, thus the refinement volumes with the smallest cell size both fall below this value and should satisfactorily resolve the TKE. The vertical refinement of the waterline around the ship was chosen such that the cell length is equal to the cell height, since \verb|interFoam| may react very sensitive to the change in gradients if the cell is not equilateral. This was realised in front of the ship, slightly before wave elevation from the bow wave starts, behind the ship, capturing the interface at least until the bilge vortex fully developed, and along the ship hull. Along the waterline anisotropic refinement was applied over the full domain, as coarsening in front of the domain borders caused an artificial rise of the water level and/ or a wave. Interestingly, this seems not to happen due to a lack of resolution, as simulations without (anisotropic) waterline refinement did not show this behaviour. Also at waterline, a generous Kelvin pattern was used for horizontal refinement around the ship. The refinement box of second-smallest cell size captures the full underwater aft ship, starting approximately where the strong curvature of the stern frames begins. The refinement box of smallest cell size is fully capturing the area where the bilge vortex develops (EFD measuring planes) and immediately behind.

The first cell size is established to $\qty{0.0015}{m}$, corresponding to $y^+ = 40$. This allows for the medium mesh to have at least two boundary layers, while for the fine mesh no layers are added within the smallest-cell-size refinement zone. Often discussions rise from this topic, but actually equilateral (quadrate) cells in the boundary layer can be an advantage in LES simulations such as the application presented here. Elongated cells in the boundary layer can be applied due to having almost exclusively tangential stresses close to the wall, at least in mainly one-directional flows, which allows for saving a lot of grid points. The meshes are unstructured fully-hexahedral grids which were created using \textit{HEXPRESS}\texttrademark . Both set-ups have a symmetry plane at centerline, the 1P meshes one at still-waterline level too. The grid size ranges between 4.3$\cdot$ $10^6$ (1P medium) and 29.7$\cdot$ $10^6$ cells (2P fine).

\begin{table}[H]
\centering
\caption{Boundary Conditions}
\label{tab:2}
\resizebox{\textwidth}{!}{%
\begin{tabular}{l|lllll}
 & inlet & outlet & top (2P) \footnotemark & \begin{tabular}[c]{@{}l@{}}bottom,\\ side,\\ midPlane\end{tabular} & JBC \\ \hline
U & fixedValue & \begin{tabular}[c]{@{}l@{}}inletOutlet (1P)\\ outletPhaseMeanVelocity (2P)\end{tabular} & pressureInletOutletVelocity & symmetry & movingWallVelocity \\
p (1P) & zeroGradient & fixedValue & symmetry & symmetry & zeroGradient \\
p\_rgh (2P) & fixedFluxPressure & zeroGradient & totalPressure & symmetry & fixedFluxPressure \\
k & fixedValue & inletOutlet & inletOutlet & symmetry & kqRWallFunction \\
$\omega$ & fixedValue & inletOutlet & inletOutlet & symmetry & omegaWallFunction \\
nut & calculated & zeroGradient & zeroGradient & symmetry & nutUSpaldingWallFunction \\
alpha.water (2P) & fixedValue & variableHeightFlowRate & inletOutlet & symmetry & zeroGradient
\end{tabular}%
}
\end{table}

\footnotetext{1P: symmetry for all parameters}

\FloatBarrier



\subsection{Case Set-Up}

\begin{wraptable}{l}{.6\textwidth}
\caption{Solution Parameters}
\label{tab:3}
\resizebox{.6\textwidth}{!}{
\begin{tabular}{lll}
term & parameter/ keyword & scheme/ value \\ \hline
&&\\
Time derivatives & ddtSchemes & CrankNicolson 0.9 \footnotemark \\
Gradient term & gradSchemes & linear \\
&&\\
Divergence term & \begin{tabular}[c]{@{}l@{}}div(phi,U)\\ div(rhoPhi,U)\\ div(phi,k)\\ div(phi,omega)\\ div(phi,alpha)\\ div(phirb,alpha) \end{tabular} & \begin{tabular}[c]{@{}l@{}}limitedLinear 0.1 \footnotemark \\ limitedLinearV 0.1\\ limitedLinear 0.1\\ limitedLinear 0.1\\ vanLeer\\ interfaceCompression vanLeer 1 \end{tabular} \\
&&\\
\begin{tabular}[c]{@{}l@{}}solver\\ alpha.water \end{tabular} & \begin{tabular}[c]{@{}l@{}}nAlphaCorr\\ nAlphaSubcycles\\ cAlpha\\ icAlpha\\ MULESCorr\\ nLimiterIter \end{tabular} & \begin{tabular}[c]{@{}l@{}}2\\ 0\\ 1\\ 0\\ yes\\ 3 \end{tabular} \\
&&\\
other& \begin{tabular}[c]{@{}l@{}}momemtumPredictor\\ nOuterCorrectors\\ nCorrectors\\ nNonOrthogonalCorrectors\\ turbOnFinalIterOnly \end{tabular} & \begin{tabular}[c]{@{}l@{}}no\\ 1\\ 2--3 \footnotemark \\ 1\\ no \end{tabular}
\end{tabular}%
}
\end{wraptable}

The simulations were mostly executed on \enquote{Hochleistungsrechner Karlsruhe} (HoReKa), a High-Performance-Computing (HPC) cluster \citep{KAR23}, while the rest was done on \enquote{Haumea}, the HPC cluster of the University of Rostock. All cases were computed using OpenFOAM v2112.\\
The simulations are initiated by carrying out a RANS simulation first. This already gives reliable results regarding drag coefficients and the wave elevation, and allows to evaluate the mesh quality. The simulation is stopped when convergence is reached and the latest solution is used a starting point for the hybrid calculations. At least two times the ship is completely passed by the flow, before the averaging of the fields is started ($t_A = \qty{14}{s}$). To be on the safe side, the averaging was conducted until a total simulation time of $t_S = \qty{55}{s}$. For all cases, the maximum Courant number is set to \verb|maxCo|$=0.5$, and at the interface to \verb|maxAlphaCo|$=0.3$. The applications used for RANS are a) 1P: \verb|simpleFoam| and b) 2P: \verb|interFoam| with large time-stepping (LTS). For URANS and SLH it is a) 1P: \verb|pimpleFoam| with one outer corrector, equalling the PISO algorithm, and b) 2P: \verb|interFoam| using PISO algorithm. A detailed overview of the solution parameters is given in table~\ref{tab:3}. The RANS and URANS computations make use of the k-$\omega$-SST model, while the hybrid computations are based on the SLH model.

\footnotetext[2]{starting with lower numbers (mostly 0.7), depending on stability of simulation; max. value applied no later than approx. 7s simulation time}
\footnotetext[3]{starting with higher numbers (mostly 0.3), depending on stability of simulation; min. value applied no later than approx. 7s simulation time}
\footnotetext[4]{depending on stability of simulation}

\FloatBarrier



\setlength{\textfloatsep}{5pt}

\section{Results}

First, a validation of the SLH model with experimental values will be presented, followed by a grid convergence study. Next, the horizontal and vertical course of the TKE, as done during Tokyo'15, will be displayed. While these results are unanticipated and not satisfactory, the circulation is examined and a number of additional test simulations carried out. After the latest results still do not provide the necessary insight, supplementary URANS simulations are done, and an analysis of the position of the vortex cores. To ensure better comparability, a single fixed center position is then defined. The medium mesh is identified as the problem and modified accordingly. Eventually, the final results are presented, including consideration of the integral quantity and spatial distribution.

\begin{wrapfigure}{l}{.5\textwidth}
	\centering
		\includegraphics[scale=.75]{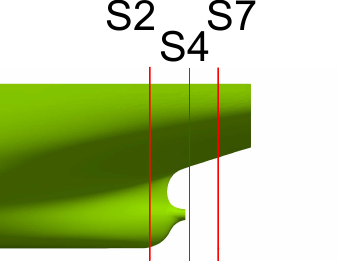}
	\caption{Cross sections S2, S4 and S7 of the JBC.}
	\label{fig:4}
\end{wrapfigure}

Resistance coefficients and the wave elevation/ wave cuts were examined with RANS in order to fully assess the quality of the grids, but are not shown for the sake of brevity. The RANS results regarding the resistance coefficient are strongly dependent on the mesh and especially the schemes. While choosing an appropriate scheme can lead to differences to experimental results less than $1\%$, less strict schemes are chosen, as RANS was simply there to provide an initial solution to SLH. The resistance coefficient of the SLH simulations is met with a maximum error of $4.03\%$ (1P) and $7.61\%$ (2P), for URANS $1.18\%$ (1P) and $4.12\%$ (2P). Further experimental results are available for seven cross sections of the ship's stern, though mainly three of them are used for evaluation (see fig.~\ref{fig:4}): S2 ($x/L_{PP} = 0.9625$), S4 between duct and propeller ($x/L_{PP} = 0.9843$), and S7 at AP ($x/L_{PP} = 1.0$).
To give a first impression on the flow features with different grids and methods, the wave contours on and next to the hull and velocity contours  are shown. The wave contours are selected to show the difference between RANS and SLH on the fine mesh with regards to the experiment, with an additional comparison of SLH on the final medium mesh. Exemplary, for the velocity contours the final improved medium mesh is chosen to show the difference between URANS and SLH and 1P/ 2P. Pictures of the 3D vortex structures based on the Q-criterion complement the insights.

\begin{figure}[!htbp]
	\centering
	\subfloat[Wave Contour on Hull]
	{\includegraphics[width=.5\linewidth]{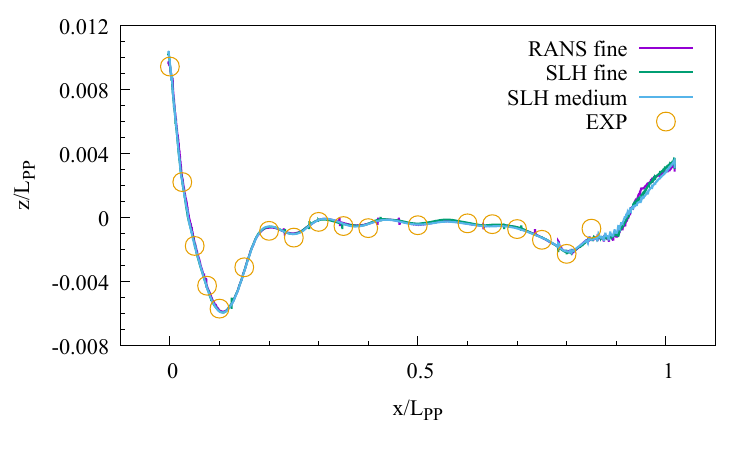}
    \label{fig:17a}}
    \subfloat[Wave Cut at $y/L_{PP} = 0.1043$]
	{\includegraphics[width=.5\linewidth]{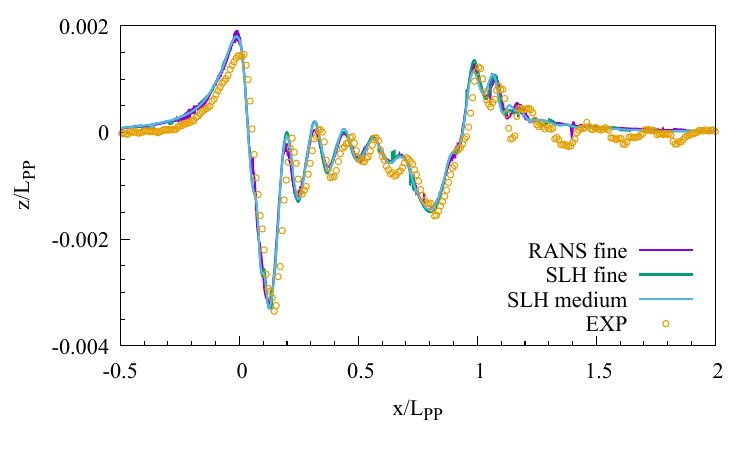}
    \label{fig:17b}}\\
    \subfloat[Wave Cut at $y/L_{PP} = 0.1900$]
	{\includegraphics[width=.5\linewidth]{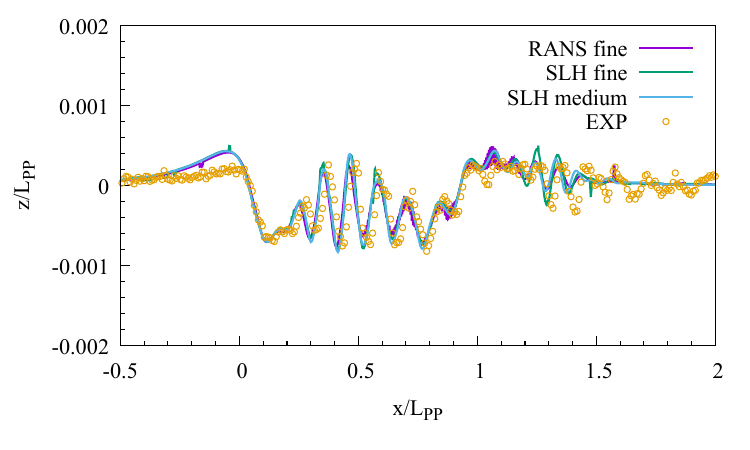}
    \label{fig:17c}}
    \caption{Comparison of wave contour and cuts for selected comparison cases.}
    \label{fig:17}
\end{figure}


The wave contour on the hull (fig.~\ref{fig:17}) is in near perfect agreement with EFD. The wave cuts next to the ship show an overestimation of the bow wave and a slight phase shift. As this behaviour was observed in all submissions to Tokyo'15, a problem with the EFD data was assumed. The deviation further away from the ship is due to mesh coarsening. The diagrams show that the different mesh sizes or usage of SLH do not significantly influence the interface.

\begin{figure}[!htbp]
	\centering
	\subfloat[1P, $Q_{inst}$ at $t = \SI{55}{s}$]
	{\includegraphics[width=.25\linewidth]{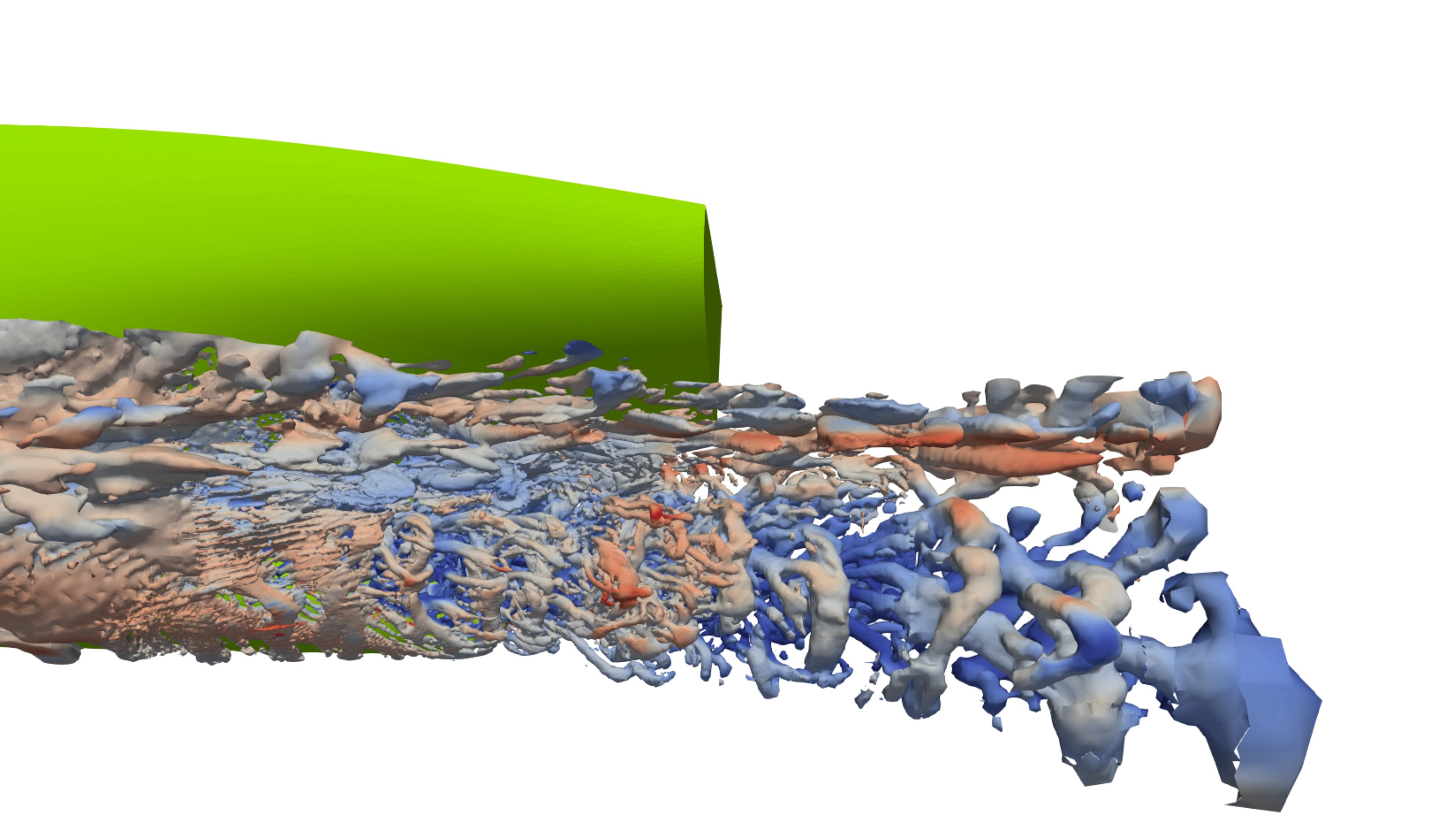}
    \label{fig:18a}}
	\subfloat[1P, $Q_{mean}$]
	{\includegraphics[width=.25\linewidth]{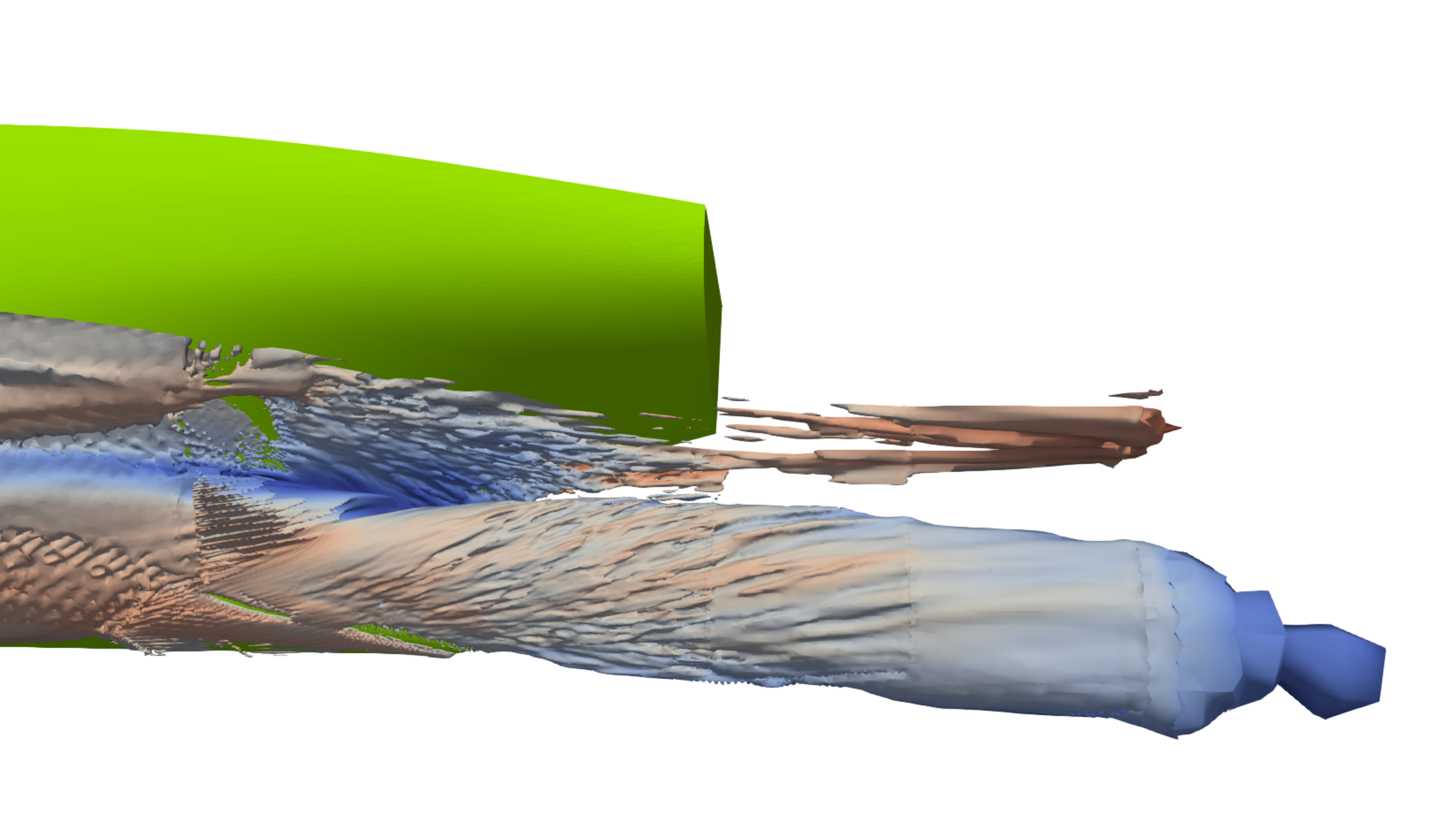}
    \label{fig:18b}}
	\subfloat[2P, $Q_{inst}$ at $t = \SI{55}{s}$]
	{\includegraphics[width=.25\linewidth]{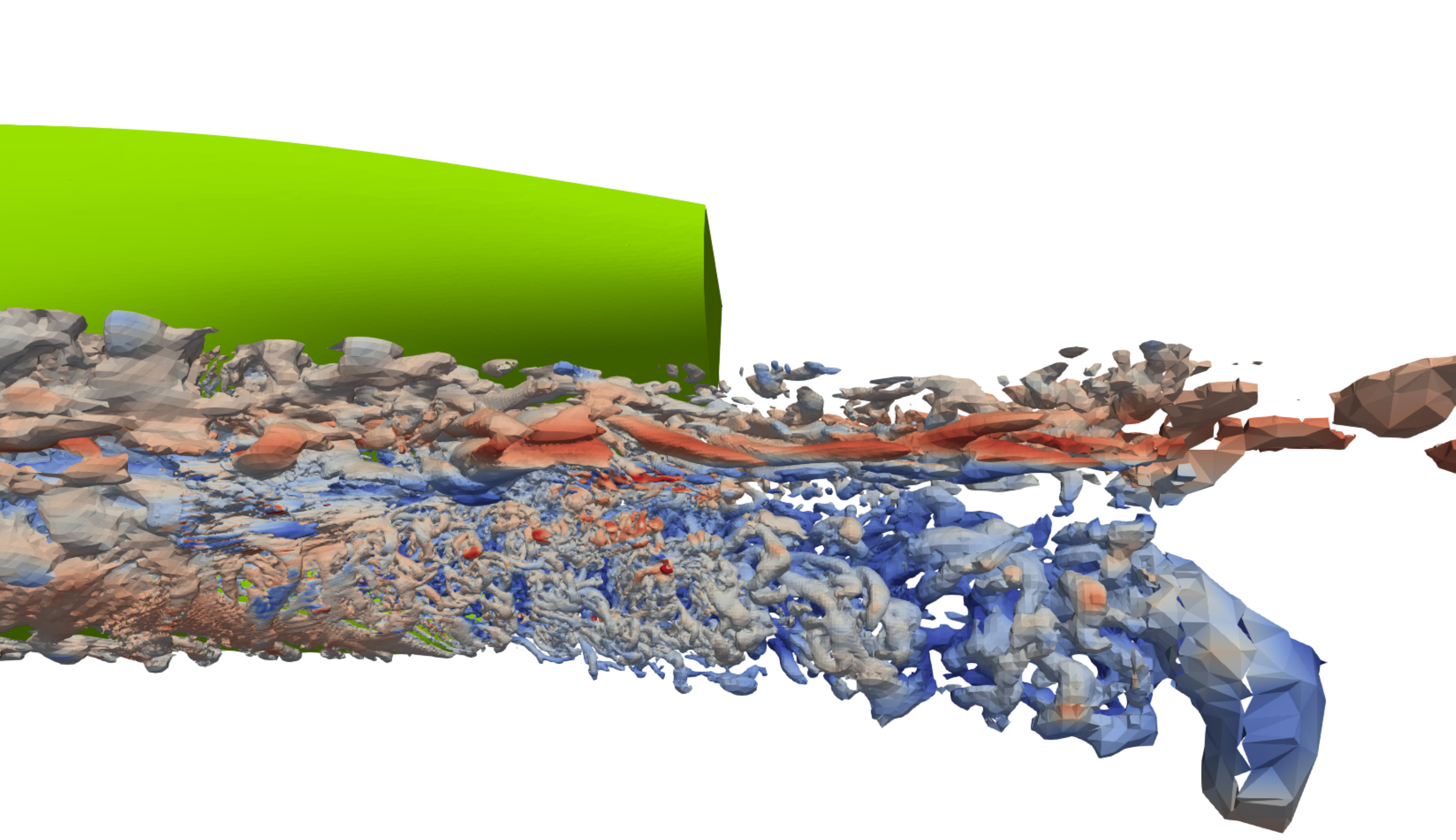}
    \label{fig:18c}}
	\subfloat[2P, $Q_{mean}$]
	{\includegraphics[width=.25\linewidth]{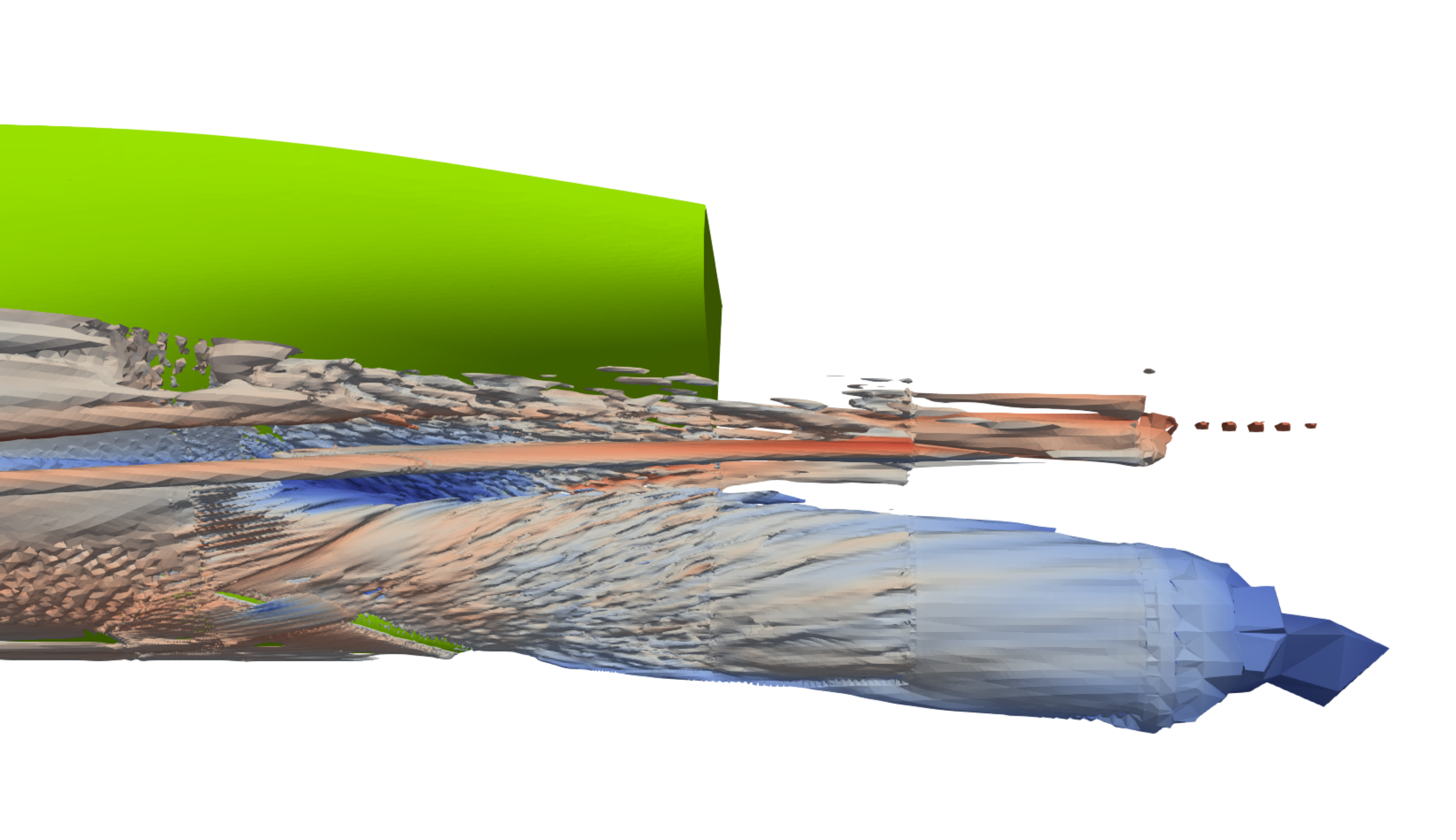}
    \label{fig:18d}}\\
	\subfloat[1P, $Q_{inst}$ at $t = \SI{55}{s}$]
	{\includegraphics[width=.25\linewidth]{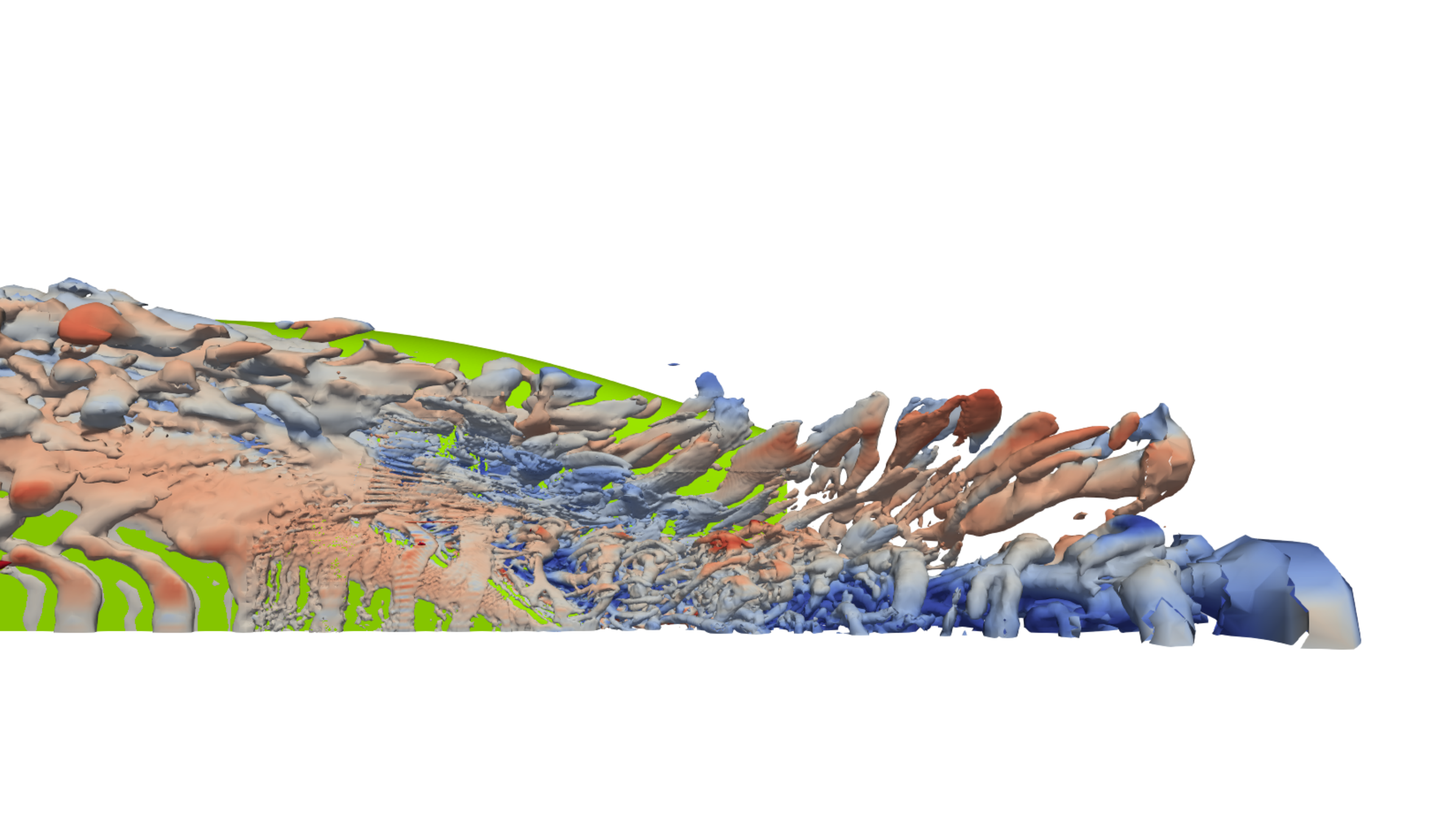}
    \label{fig:18e}}
	\subfloat[1P, $Q_{mean}$]
	{\includegraphics[width=.25\linewidth]{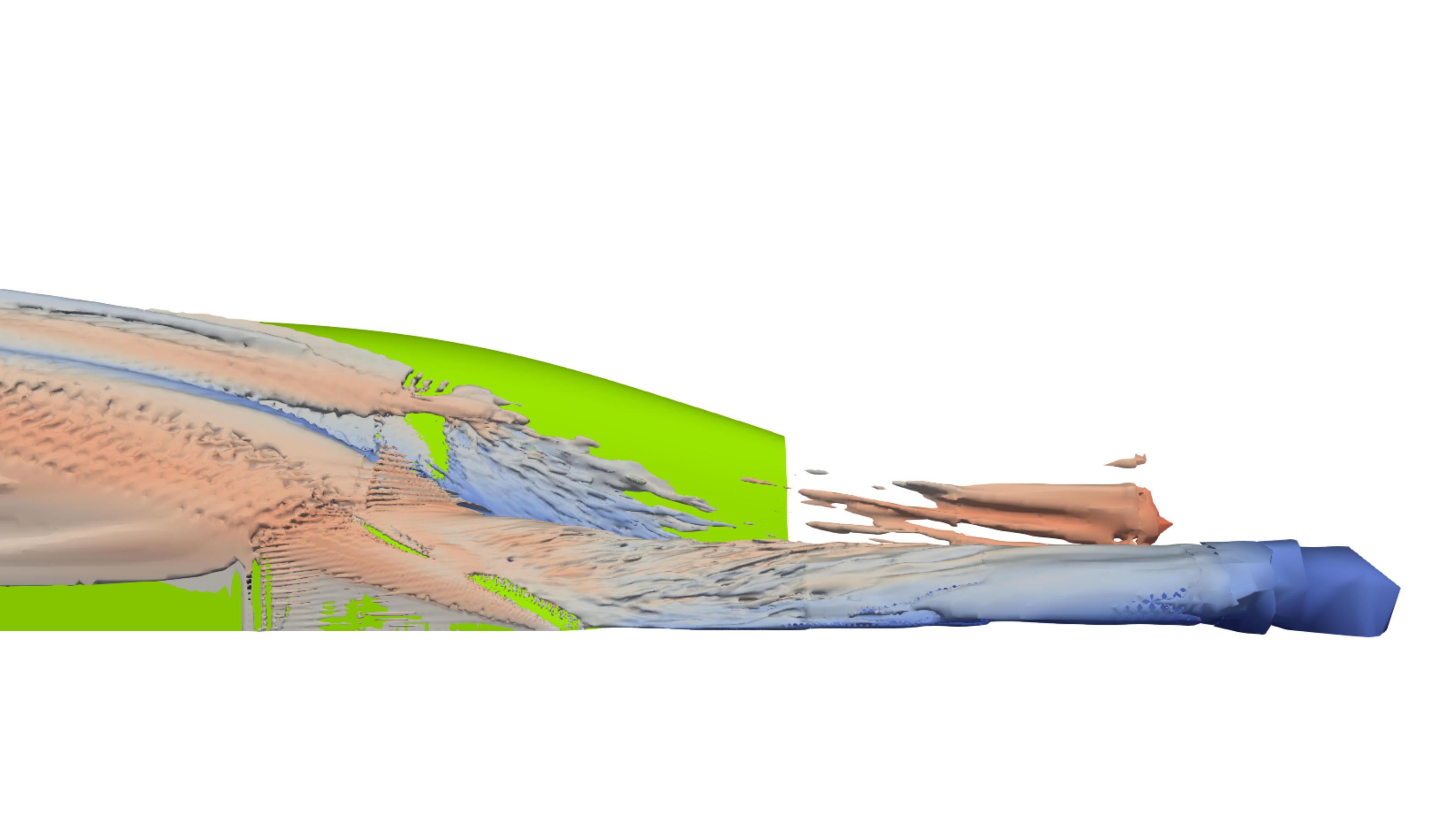}
    \label{fig:18f}}
	\subfloat[2P, $Q_{inst}$ at $t = \SI{55}{s}$]
	{\includegraphics[width=.25\linewidth]{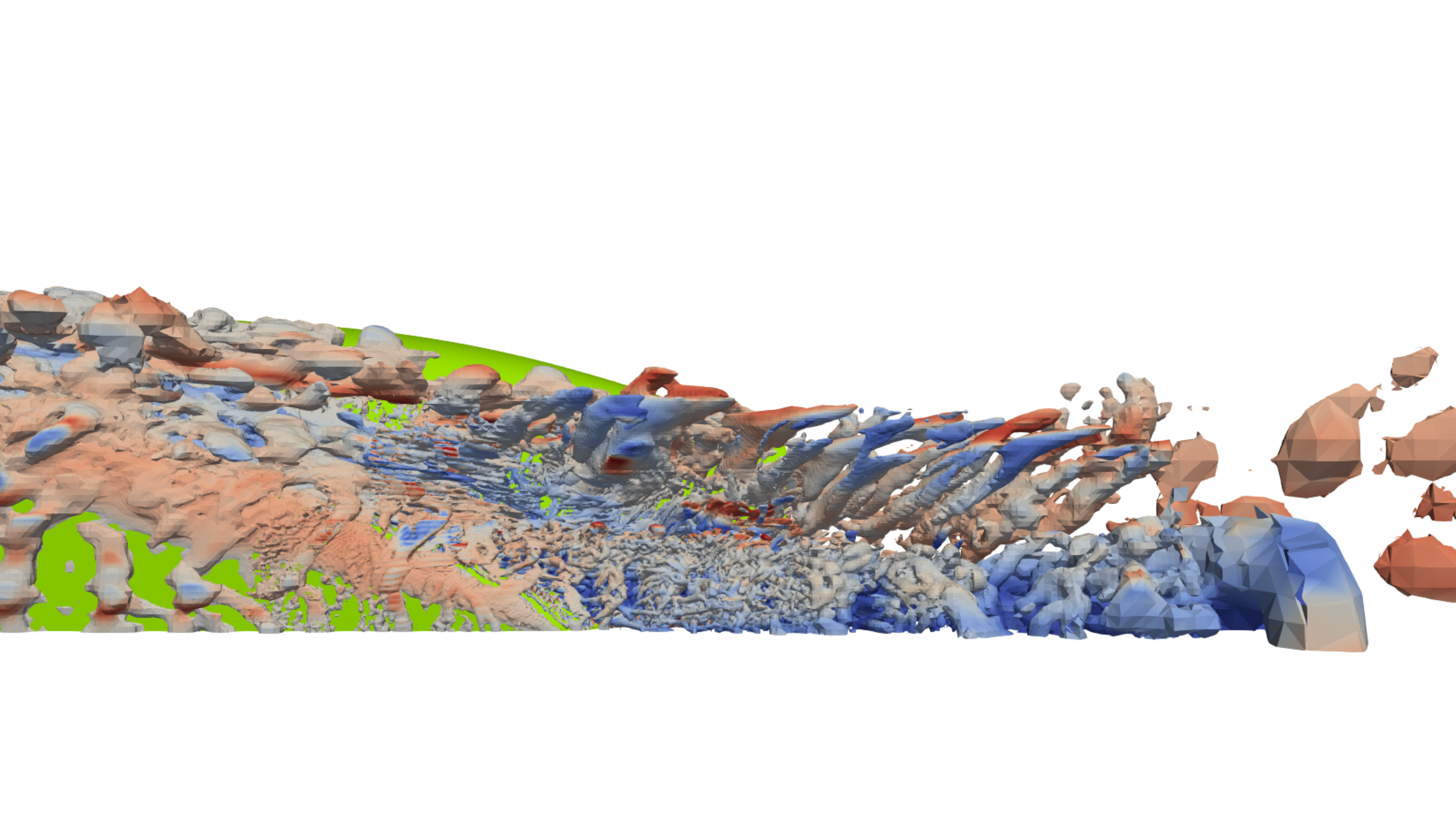}
    \label{fig:18g}}
	\subfloat[2P, $Q_{mean}$]
	{\includegraphics[width=.25\linewidth]{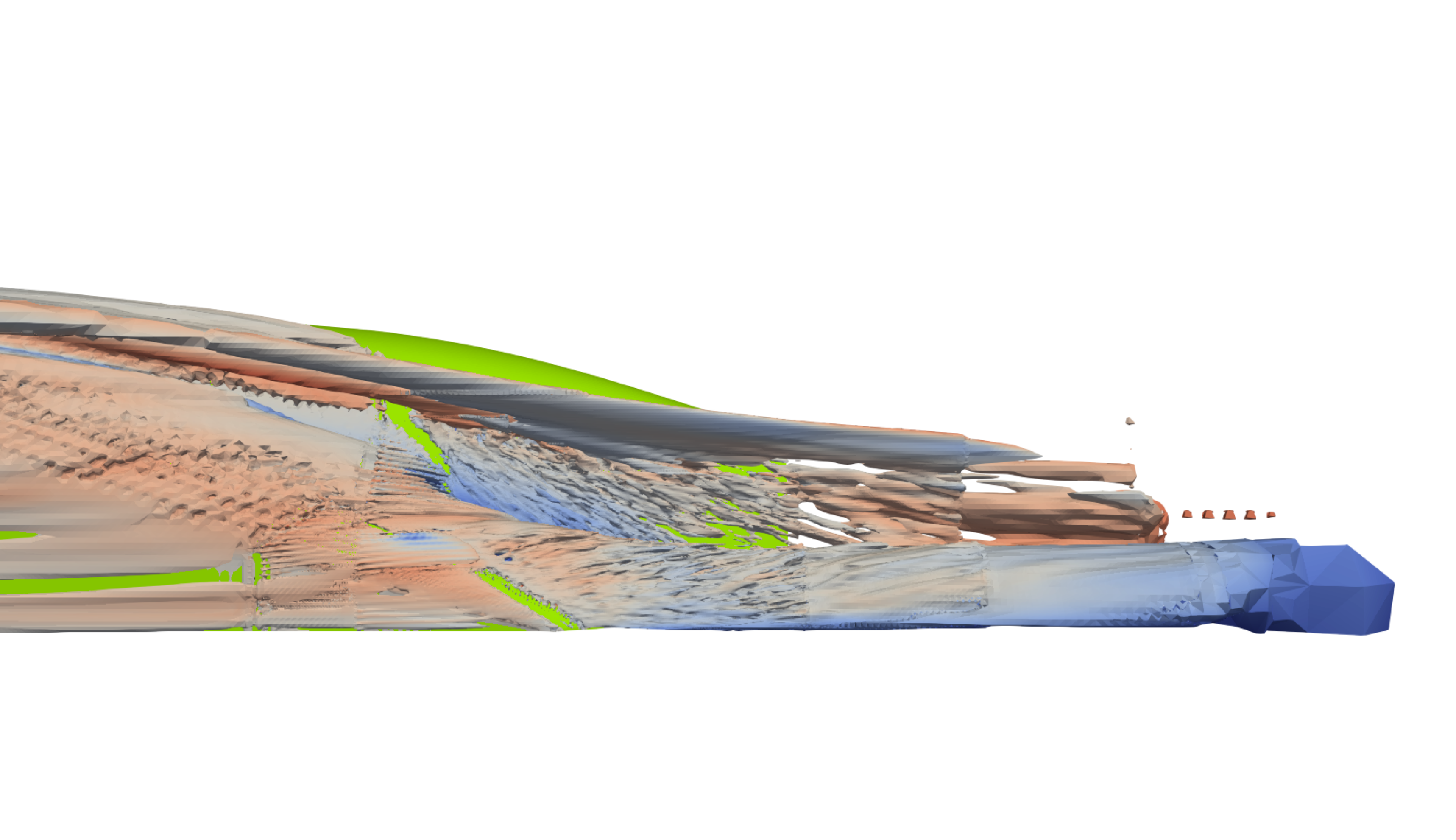}
    \label{fig:18h}}
    \caption{Contour of time-averaged and instantaneous vortical structures based on $Q^* = 25$ for side view (top) and view from below (bottom) at aft part of ship, with and without free surface, on final medium mesh.}
    \label{fig:18}
\end{figure}


According to Tokyo'15 the participants should show vortical structures using the Q-criterion $Q^*$, which is defined as $Q^* = Q \cdot{ L_{PP}}^2 / {U_{ref}}^2$ where $U_{ref}$ is the design speed specified for the model. They are coloured by helicity
\begin{equation}
	\label{eqn:helicity}
	helicity = \frac{ U_{mean} \cdot \Omega_{mean}}{|U_{mean}| \cdot |\Omega_{mean}|}
\end{equation}

The side views show the two averaged contours (figs. \ref{fig:18b}, \ref{fig:18d}) to be very alike regarding dimensions and orientation, while the helicity points to exiguous differences in spin. The 2P case depicts a larger amount of structures, certainly due to turbulence at the surface. The instantaneous contours reveal a kind of bilge vortex in its typical form, although it can only be discerned a faint distance behind the ship. Further forward, numerous small and medium-sized vertices do not reveal a single definable structure, although they evidently all follow the same pattern of movement.

\begin{figure}[!htbp]
	\centering
	\vspace{-1.75em}
	\subfloat[EFD]
	{\includegraphics[width=.33\linewidth]{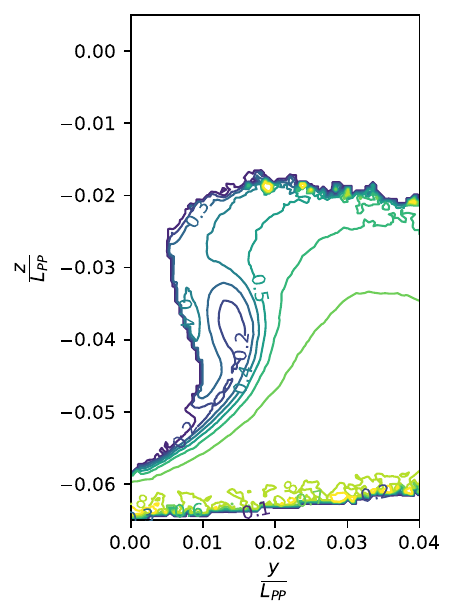}
    \label{fig:16a}}
    \subfloat[SLH (1P)]
	{\includegraphics[width=.33\linewidth]{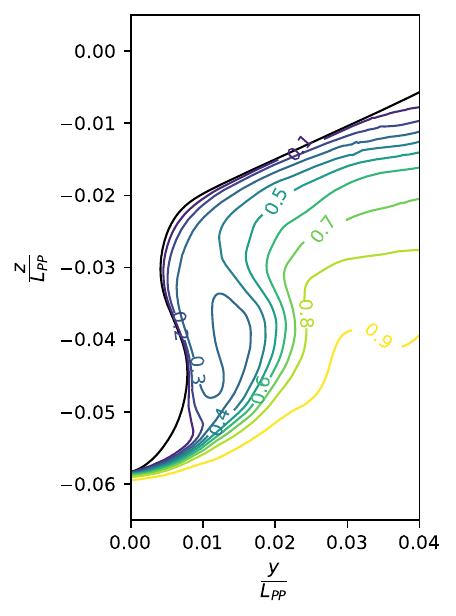}
    \label{fig:16b}}
    \subfloat[SLH (2P)]
	{\includegraphics[width=.33\linewidth]{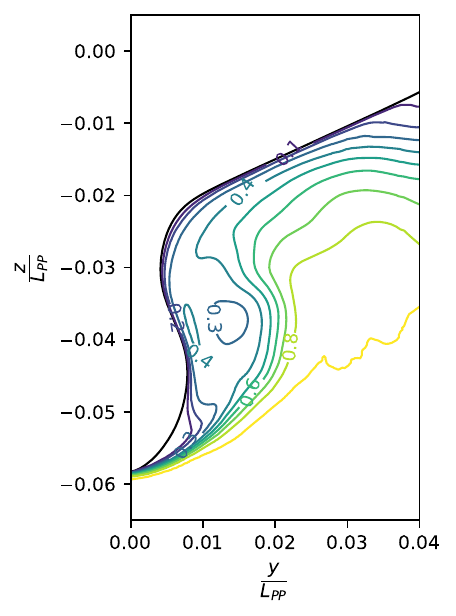}
    \label{fig:16c}}
    \phantomcaption
    \vspace{-1.25em}
    \end{figure}
    \begin{figure}[!htbp]\ContinuedFloat
    \centering
    \hfill
    \subfloat[URANS (1P)]
	{\includegraphics[width=.33\linewidth]{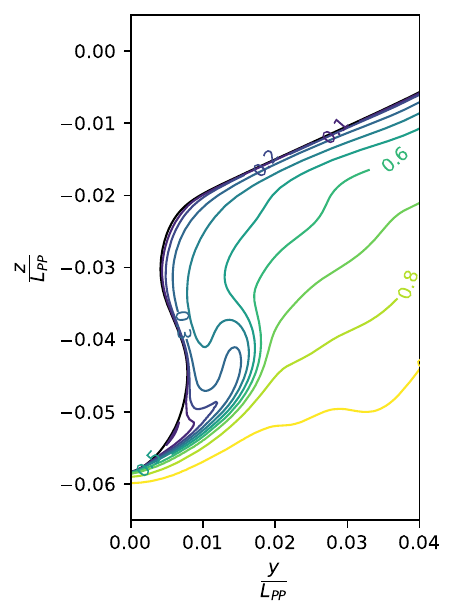}
    \label{fig:16d}}
    \subfloat[URANS (2P)]
	{\includegraphics[width=.33\linewidth]{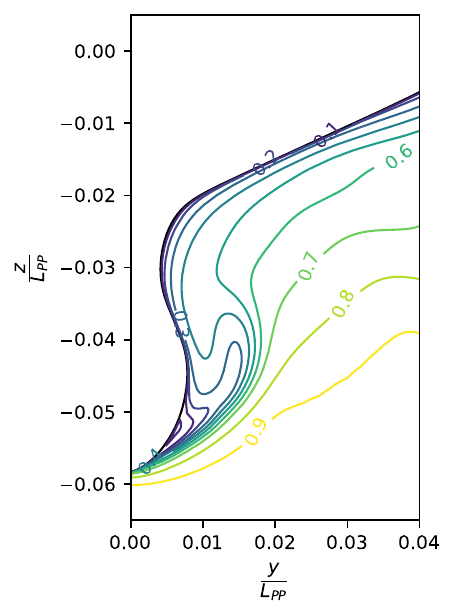}
    \label{fig:16e}}
    \caption{$U_{xx}$ at cross section S2 on final medium mesh.}
    \label{fig:16}
\end{figure}

\FloatBarrier

The diagrams (fig.~\ref{fig:16}) visualise the velocity contours in the wake of the JBC. All meshes and methods show the typical hook shape, whereby for URANS it is not well-developed and the velocity too high. Both hybrid simulations can reproduce the EFD shape, while the velocity in the center is slightly overpredicted. While the hybrid single-phase mesh meets the innermost contour better, the two-phase mesh identifies the small isowake contour close to the duct. The first measurements by NMRI showed the innermost contour as an isowake island, while the latest measurements show a nearly continuous shape with a thin connection to the lower part. However, the results are sufficiently similar. As the distance to the duct increases, the degree of correspondence is slightly reduced. This is in accordance with findings of Tokyo'15 regarding the underlying k-$\omega$-SST model.

\subsection{Grid Convergence}
\label{sec:gridConv}

\begin{figure}[!htbp]
	\centering
	\subfloat[Fraction of TKE\textsubscript{res} on \textbf{coarse mesh}]
	{\includegraphics[width=.33\linewidth]{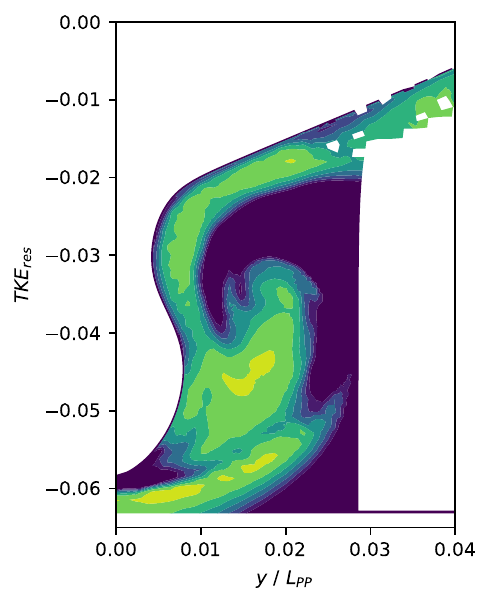}
    \label{fig:7d}}
    \subfloat[Fraction of TKE\textsubscript{res} on \textbf{medium mesh}]
	{\includegraphics[width=.33\linewidth]{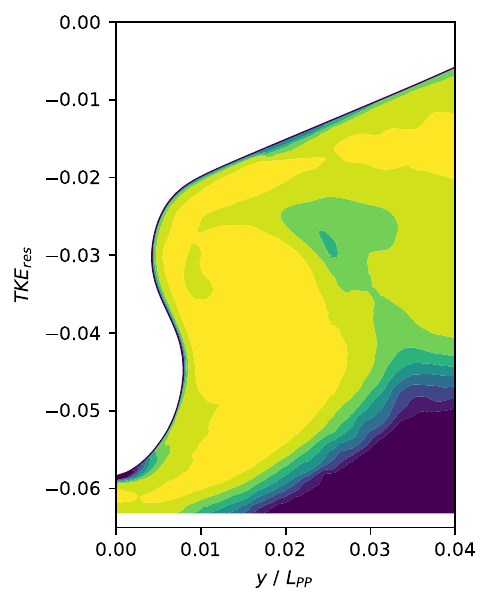}
    \label{fig:7e}}
    \subfloat[Fraction of TKE\textsubscript{res} on \textbf{fine mesh}]
	{\includegraphics[width=.33\linewidth]{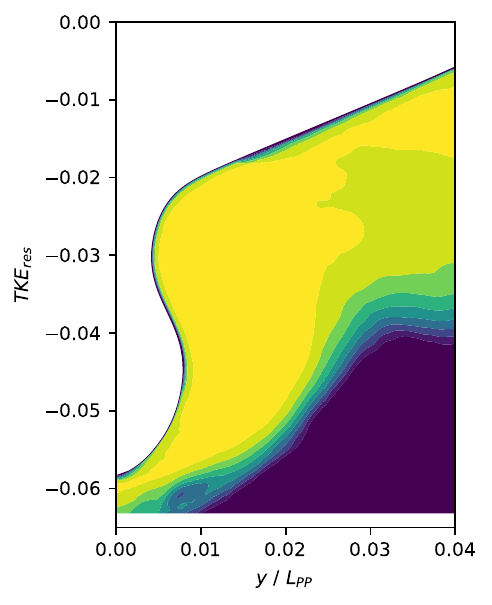}
    \label{fig:7f}}\\
    \subfloat
	{\includegraphics[scale=.8]{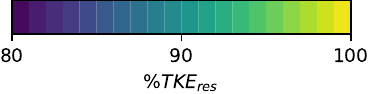}}
    \caption{Share of resolved TKE for \textbf{2P} meshes at \textbf{S2}.}
    \label{fig:7}
\end{figure}

In CFD it is mandatory to present grid-convergence studies. While this makes sense for the majority of cases, it gets complicated when using hybrid methods. Of course, there are some tendencies which are expected when the mesh gets refined, but on the other hand one should keep in mind the uncertainty of the used methods. Often, convergence studies are done with regards to forces coefficients and hybrid methods are expected to have an error around $10...15\%$, at least for the usual "control parameter", the drag coefficient. That said, it is impossible to differentiate, if discrepancies less than this number are due to the mesh or the method. Hence, other criteria are chosen to demonstrate the method to work as intended and show the differences between the mesh sizes.\\
First, the share of URANS/ LES is tracked to illustrate the switching. Second, the portion of resolved TKE is determined. If at least $80\%$ are resolved, it is considered successful \citep[p. 241]{POP00}. With increasing mesh size the expectation is to have a larger share of resolved TKE.\\
Exemplary, three 2P meshes with different grid size are shown at cross section S2 (fig.~\ref{fig:7}).
It shows, that following the expectations the share of resolved TKE is increasing with smaller cell size. For the \enquote{medium} mesh already a share of nearly $100\%$ is accomplished. While for the fine mesh the overall area of values above $80\%$ is a little smaller compared with the medium mesh, the area of interest is fully covered with even higher values.

\subsection{Results following guidelines of Tokyo'15 workshop}
\label{sec:resultA}

For comparison purposes, initially it is followed the guidelines on depiction of TKE of the Tokyo'15 workshop. Hence, the procedure to determine the vortex center ($y_v$, $z_v$) is finding $\Omega_{x,max}$ and from there a horizontal and vertical line with the length $(y - y_v)/L_{PP} = 0.005$ in each direction gathers the TKE values (fig.~\ref{fig:5}). Details are available on the Tokyo'15 website \citep{NAT15b}. To stay consistent, the same axis dimensions are chosen as proposed for the workshop. This will make it easier to compare with data of other researchers.

\subsubsection{TKE}
\label{sec:A_tke}

The portrayed lines consist of at least 100 points, highlighting every tenth point. Increasing the number of points up to 1000 did not reveal any outliers.

\begin{figure}[!htbp]
	\centering
	\subfloat[Horizontal distribution of TKE at \textbf{S2}.]
	{\includegraphics[scale=1]{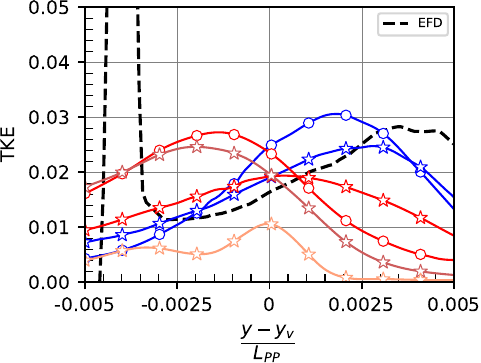}
    \label{fig:5a}}
    \subfloat[Vertical distribution of TKE at \textbf{S2}.]
	{\includegraphics[scale=1]{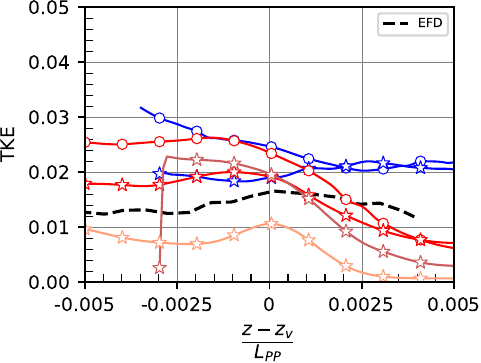}
    \label{fig:5d}}
    \phantomcaption
    \end{figure}
    \begin{figure}[!htbp]\ContinuedFloat
    \centering
    \subfloat[Horizontal distribution of TKE at \textbf{S4}.]
	{\includegraphics[scale=1]{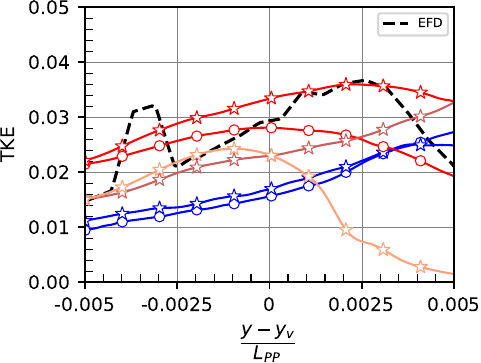}
    \label{fig:5b}}
    \subfloat[Vertical distribution of TKE at \textbf{S4}.]
	{\includegraphics[scale=1]{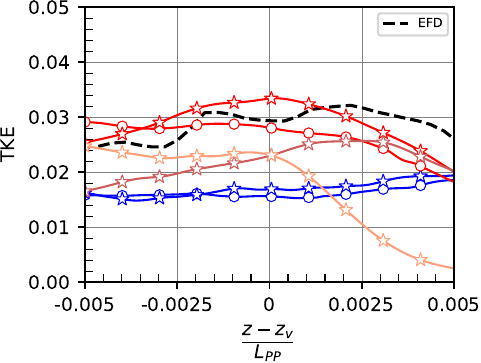}
    \label{fig:5e}}\\
    \subfloat[Horizontal distribution of TKE at \textbf{S7}.]
	{\includegraphics[scale=1]{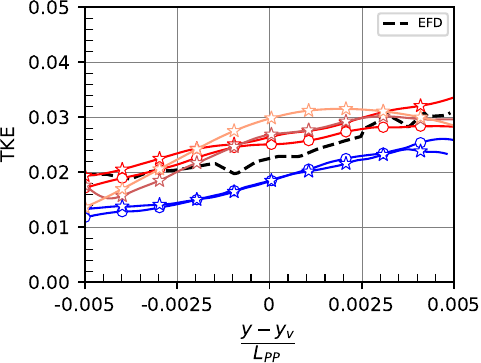}
    \label{fig:5c}}
    \subfloat[Vertical distribution of TKE at \textbf{S7}.]
	{\includegraphics[scale=1]{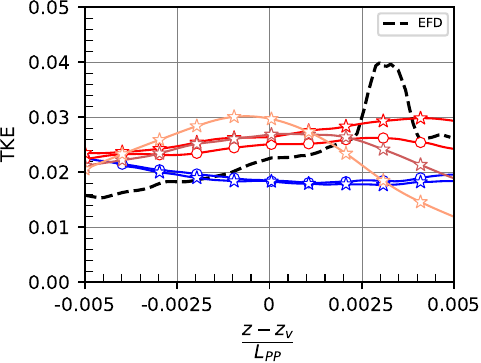}
    \label{fig:5f}}\\
    \subfloat
	{\includegraphics[scale=0.7]{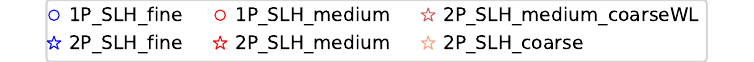}}
    \caption{Distribution of TKE at S2, S4 and S7.}
    \label{fig:5}
\end{figure}

\FloatBarrier

The depicted results show a rather inconsistent behaviour. For S2 the TKE is overestimated, respectively the maximum is shifted (apart from the coarse mesh). In general it can be seen that the TKE increases with increasing mesh refinement, apart from the finest mesh. One exception is the coarse mesh at S7, most likely as the vortex structure is differently developed in comparison. While at S2 the single-phase TKE is significantly greater than for the two-phase simulation, this turns into the opposite for S4. At S7 the two-phase TKE is still bigger for the medium mesh, while the results of the fine meshes for 1P and 2P are highly alike. Apart from S2, which seem to meet the experimental results pretty well, the fine meshes show a significantly (up to 50$\%$) lower amount of TKE than the medium meshes.

\subsubsection{Circulation}
\label{sec:A_circ}

To get a more reliable assertion about the vortex structure, the circulation starting from the respective determined vortex center is calculated. The expectation is, that as long as the integration contour is within the structure, the value is increasing. Covering at least the range where the course of the TKE is illustrated (see fig.~\ref{fig:4}), a steady increase points to expected behaviour. Outside of the main structure, an asymptotic behaviour is expected \citep{DEV96}. For better clarity, the figures shown here do not extend to this range. However, they were produced and show the anticipated asymptotic drop.\\
Circulation is a measure of rotation on a closed contour and defined as \citep[p. 109]{NEW17}:
\begin{equation}
	\label{eq:circ}
	\Gamma = \iint_S \Omega \,dS
\end{equation}

with $\,dS = n\cdot\,ds$ and $\,ds = r\cdot\,d\alpha$ where $r$ is the radius of the closed contour. Circles are created which evenly cover approximately the region of the main vortex, starting from the individual centers, and the circulation is subsequently summed up with increasing radius. If parts of the circles are outside of the fluid (i.e. inside the hull contour), this partial result is excluded and the integral limit is adjusted accordingly.

\begin{figure}[!htbp]
	\centering
	\subfloat[Circulation at \textbf{S2} from respective vortex core along different radii]
	{\includegraphics[scale=1]{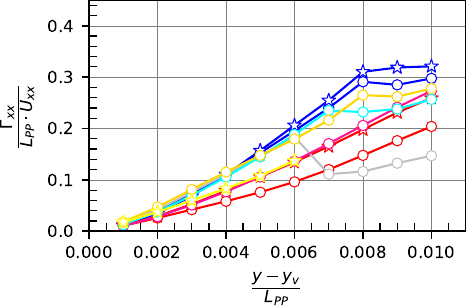}
    \label{fig:6a}}
    \subfloat[Circulation at \textbf{S4} from respective vortex core along different radii]
	{\includegraphics[scale=1]{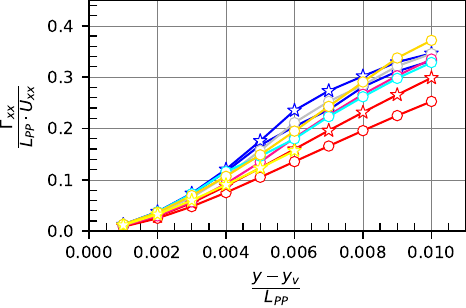}
    \label{fig:6b}}\vfill
    \subfloat[Circulation at \textbf{S7} from respective vortex core along different radii]
	{\includegraphics[scale=1]{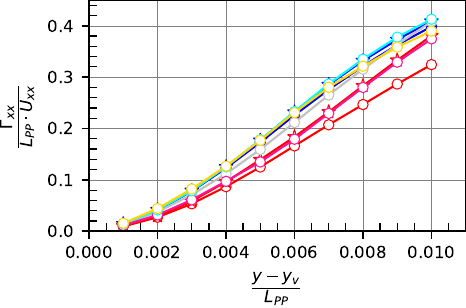}
    \label{fig:6c}}\\
    \subfloat
	{\includegraphics[scale=.7]{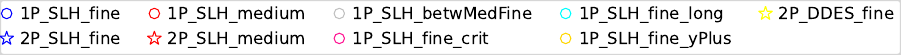}}
    \caption{Circulation at S2, S4 and S7.}
    \label{fig:6}
\end{figure}

\FloatBarrier

The diagrams (fig.~\ref{fig:6}) pretty much meet the expectations: with decreasing mesh size $\Delta\overrightarrow{x}$, circulation grows, and the tendency is 2P being greater than 1P results. However, there are some bumps and unexpected behaviour with growing radius. That is why a closer look at the vortex structure is taken, where it can be seen, that outside of the vortex structure and close to the hull (separation layer), the axial vorticity becomes tremendously negative. As the hull structure changes with the cross sections, as well as the points of contact are different due to the sundry center coordinates, the results are influenced diversely. Especially for S2 it becomes clear, that nearly half of the circles is outside of the vortex structure, and even the fluid itself. Thus, the diagrams were updated to show only the positive or the negative parts of circulation (not shown in this paper). While the pictures and the diagrams show why the lines behave that way with increasing radius, still the purely positive graphs do not proceed as thought. On the one hand, the tendency between the mesh sizes is no longer correct, on the other hand the relation between 1P and 2P turns into the opposite.

\subsubsection{Analysis of additional results}
\label{sec:A_findings}

\begin{figure}[!ht]
	\centering
	\subfloat[Horizontal distribution of TKE at \textbf{S2}.]
	{\includegraphics[scale=1]{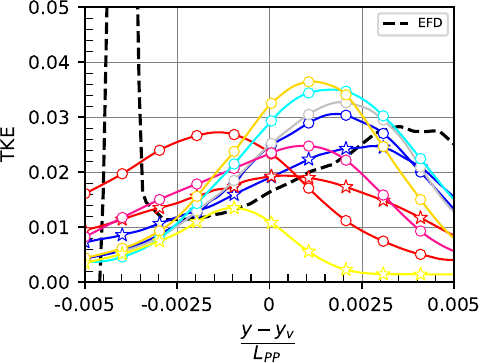}
    \label{fig:8a}}
    \subfloat[Horizontal distribution of TKE at \textbf{S4}.]
	{\includegraphics[scale=1]{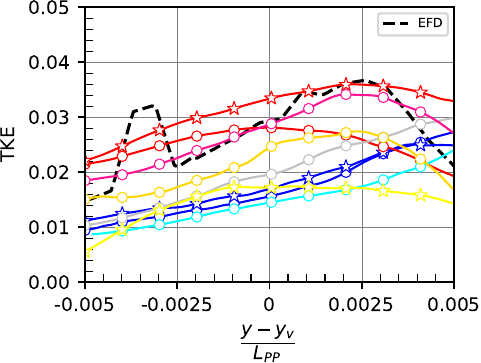}
    \label{fig:8b}}\vfill
    \subfloat[Horizontal distribution of TKE at \textbf{S7}.]
	{\includegraphics[scale=1]{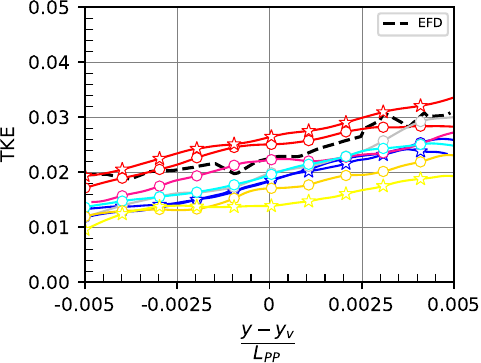}
    \label{fig:8c}}\\
    \subfloat
	{\includegraphics[scale=0.7]{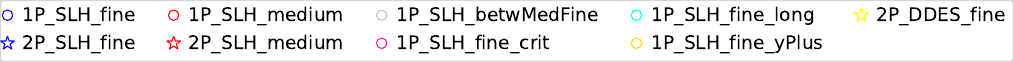}}
    \caption{Distribution of TKE at S2, S4 and S7 with test simulations.}
    \label{fig:8}
\end{figure}

Both the circulation and resolved TKE show increasing values with larger grid point number. This is expected and should not be the case if the SLH method had a serious limitation, or GIS occurred. Unfortunately it does not explain why the fine mesh shows such a huge deviation. As mentioned in section \ref{sec:dis_1_TKE}, some more simulations are carried out, to proof the point. Also, more cross sections, especially in front of S2, are analysed. The idea is, that maybe a flow characteristics further forward is not correctly detected by one of the mesh configurations, and thus the missing information produces wrong results.\\
All the tests are done on the fine mesh, and except for DDES only for 1P to save time and resources. The figures (fig.~\ref{fig:8}) show again the TKE, now with the test simulations, exemplary only the horizontal course (vertical course shows similar tendencies).

Interestingly the results do not provide the assumed behaviour. The simulation where the criterion is doubled is now closest to EFD and the good medium-mesh results compared with the rest. The other test simulations are all much closer to the fine-mesh results than of the medium mesh.

\subsection{URANS comparison and change in reference point}
\label{sec:resultB}

\subsubsection{URANS according to Tokyo'15 guidelines}

Following, results of TKE with URANS (compared with SLH results), following the Tokyo'15 procedure, are presented in figure~\ref{fig:9}:

\begin{figure}[!htbp]
	\centering
	\subfloat[Horizontal distribution of TKE at \textbf{S2}.]
	{\includegraphics[scale=1]{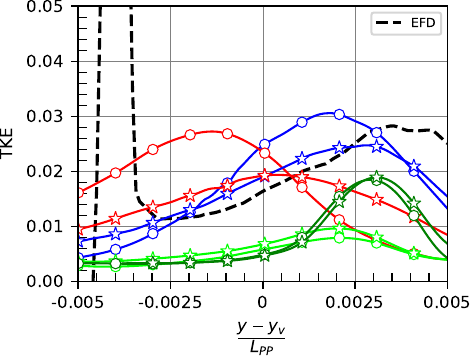}
    \label{fig:9a}}
    \subfloat[Horizontal distribution of TKE at \textbf{S4}.]
	{\includegraphics[scale=1]{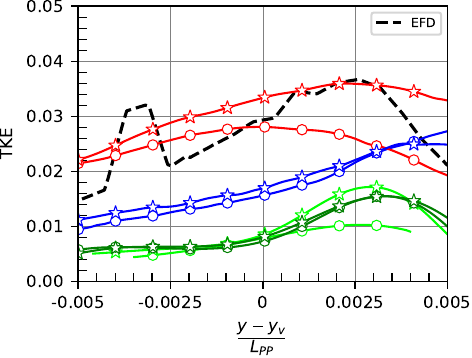}
    \label{fig:9b}}
    \phantomcaption
    \end{figure}
    \begin{figure}[!htbp]\ContinuedFloat
    \centering
    \subfloat[Horizontal distribution of TKE at \textbf{S7}.]
	{\includegraphics[scale=1]{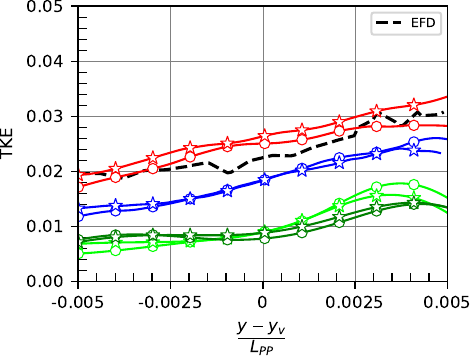}
    \label{fig:9c}}\\
    \subfloat
	{\includegraphics[scale=0.8]{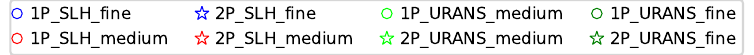}}
    \caption{Distribution of TKE at S2, S4 and S7 with test simulations (URANS).}
    \label{fig:9}
\end{figure}

\FloatBarrier

First, it should be noted that URANS finds the peaks reasonably well, whereas the magnitude is too low. What is striking is that especially at S2 and S4 the course of the graphs shows huge differences.

\subsubsection{Fixed Position}

As the different origins are not considered to be comparable at all (see \ref{sec:dis_1_vortCore}), it is decided to choose a unique point where all future observations start from, namely the vortex center of the fine single-phase mesh. Additionally, the range is extended to see what is happening outside of the presumed vortex center. It is important to point out that, unlike the vortex cores, the general structure of TKE is similar between the different cases. Hence, choosing a fixed point will not change the global behaviour or e.g. lie outside of the core. The expectation is to get a much more homogeneous course.

\begin{figure}[!htbp]
	\centering
	\subfloat[Horizontal distribution of TKE at \textbf{S2}.]
	{\includegraphics[scale=1]{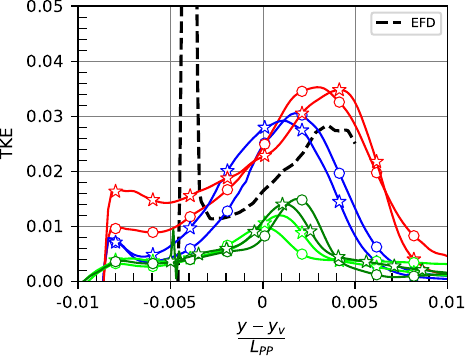}
    \label{fig:11a}}
    \subfloat[Horizontal distribution of TKE at \textbf{S4}.]
	{\includegraphics[scale=1]{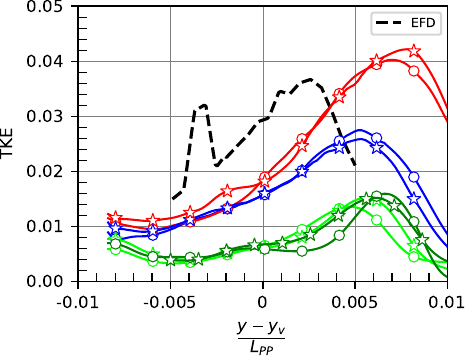}
    \label{fig:11b}}\vfill
    \subfloat[Horizontal distribution of TKE at \textbf{S7}.]
	{\includegraphics[scale=1]{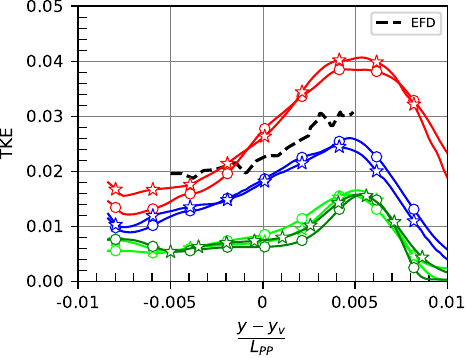}
    \label{fig:11c}}\\
    \subfloat
	{\includegraphics[scale=0.8]{legend_tkediagram_02.pdf}}
    \caption{Distribution of TKE at S2, S4 and S7 with fixed center coordinate.}
    \label{fig:11}
\end{figure}

\FloatBarrier

It can be seen, that the graphs (fig.~\ref{fig:11}) are much closer to each other. The big differences between 1P and 2P vanished, especially with regards to the grid-size uncertainty observable from the URANS graphs. At S2 and S4 the results of the medium mesh are much closer to the ones of the fine mesh, no longer being much higher, and thus closer to the EFD. At S7 the medium mesh produces results higher than for the fine mesh, but slightly off from the EFD course. Overall the results of the fine mesh are closest to EFD. Thanks to the extended range, a spatial shift is clearly visible. The pure course meets EFD but is shifted to the left. This can be explained by having a look at the position of the vortex cores (fig.~\ref{fig:10}), where at S2 the EFD coordinate is further on the left side, and at S4 a little on the right.

\subsection{Final results and integral TKE}
\label{sec:resultD}

\subsubsection{TKE from fixed position}

With the new medium meshes simulations are carried out and the procedures as described before is followed. Starting from finding the vortex cores, the picture is now different, as the distribution of vorticity, respectively the position of the maximum, is now much more alike to the fine meshes. Hence, also the positions of the vortex cores are now considerably closer together, with the medium-mesh centers overlapping the ones of the fine mesh. However, to stay consistent and ensure comparability, all lines are created from the same starting point, again the vortex center of the 1P fine grid.

\begin{figure}[!htbp]
	\centering
	\subfloat[Horizontal distribution of TKE at \textbf{S2}]
	{\includegraphics[width=.5\linewidth]{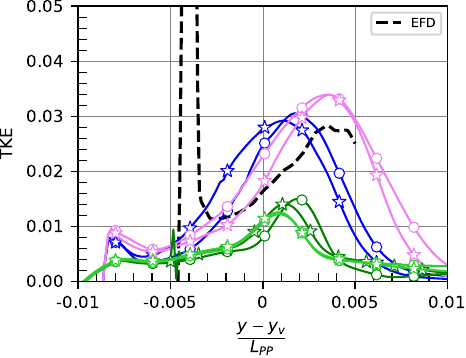}
    \label{fig:13a}}
	\subfloat[Horizontal distribution of TKE at \textbf{S4}]
	{\includegraphics[width=.5\linewidth]{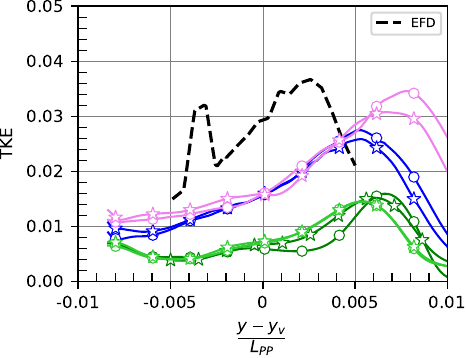}
    \label{fig:13b}}\\
    \subfloat[Horizontal distribution of TKE at \textbf{S7}]
	{\includegraphics[width=.5\linewidth]{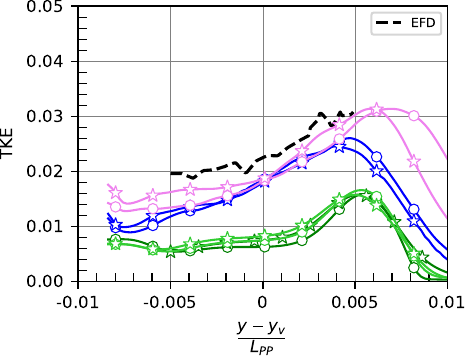}
    \label{fig:13c}}\\
    \subfloat
	{\includegraphics[scale=.8]{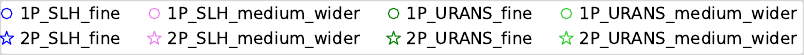}}
    \caption{Distribution of TKE at S2, S4 and S7 with new medium mesh from fixed position.}
    \label{fig:13}
\end{figure}


Apart from S2 the lines (fig.~\ref{fig:13}) are now much closer to each other (URANS/ SLH respectively). Inside the suppositious eddy center $-0.005 \leq \frac{y - y_v}{L_{PP}} \geq 0.005$ the meshes are overlapping for the most part. The difference in TKE between the mesh sizes is bigger than the differences between 1P and 2P overall and no unequivocal tendency regarding a correlation between 1P and 2P simulations is ascertainable.

\subsubsection{Integral amount of TKE}

\begin{figure}[!htbp]
	\centering
	\subfloat[Integral TKE at \textbf{S2}]
	{\includegraphics[width=.5\linewidth]{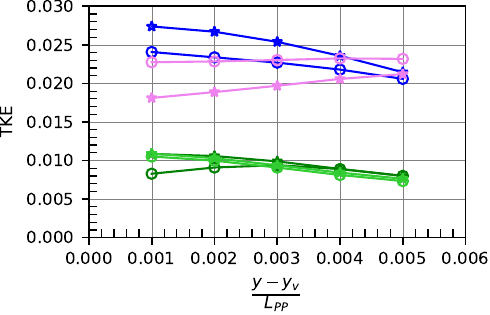}
    \label{fig:14a}}
	\subfloat[Integral TKE at \textbf{S4}]
	{\includegraphics[width=.5\linewidth]{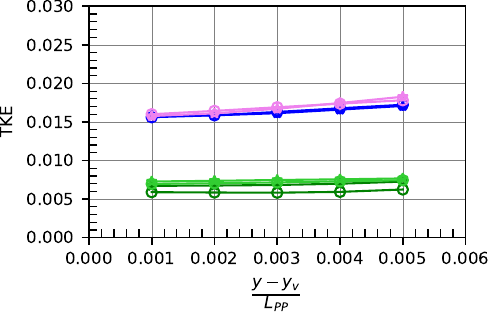}
    \label{fig:14b}}\\
    \subfloat[Integral TKE at \textbf{S7}]
	{\includegraphics[width=.5\linewidth]{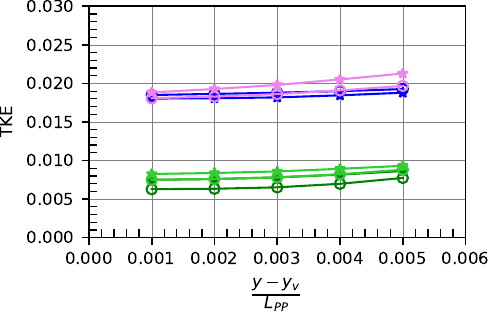}
    \label{fig:14c}}\\
    \subfloat
	{\includegraphics[scale=.8]{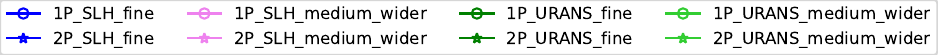}}
    \caption{Integral TKE at different cross sections.}
    \label{fig:14}
\end{figure}

Apart from S2, the results for the different mesh sizes lie very close to each other (fig.~\ref{fig:14}). It is evident from the URANS results, that differences between 1P and 2P are within the error margin of the distinct grid sizes. Actually, the only remarkable difference occurs at S2.

\subsubsection{Distribution of TKE}
\label{sec:D_distrTKE}

\begin{figure}[!htbp]
	\centering
	\subfloat[\textbf{1P} at \textbf{S2}]
	{\includegraphics[width=.33\linewidth]{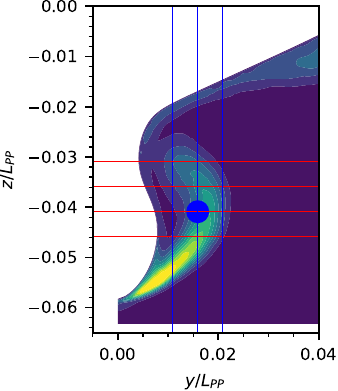}
    \label{fig:15a}}
	\subfloat[\textbf{1P} at \textbf{S4}]
	{\includegraphics[width=.33\linewidth]{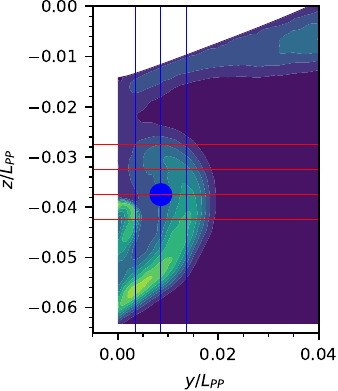}
    \label{fig:15b}}
    \subfloat[\textbf{1P} at \textbf{S7}]
	{\includegraphics[width=.33\linewidth]{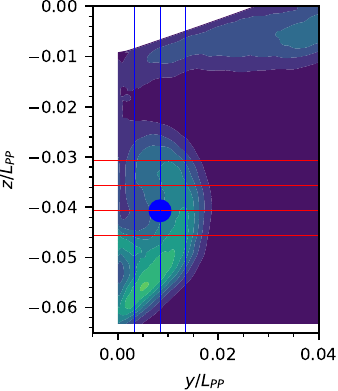}
    \label{fig:15c}}\\
    \subfloat[\textbf{2P} at \textbf{S2}]
	{\includegraphics[width=.33\linewidth]{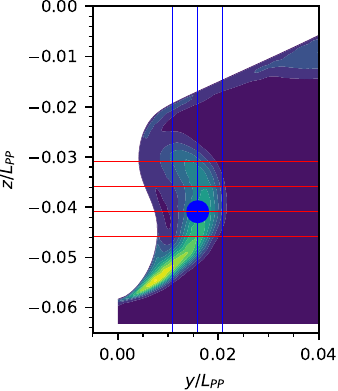}
    \label{fig:15d}}
	\subfloat[\textbf{2P} at \textbf{S4}]
	{\includegraphics[width=.33\linewidth]{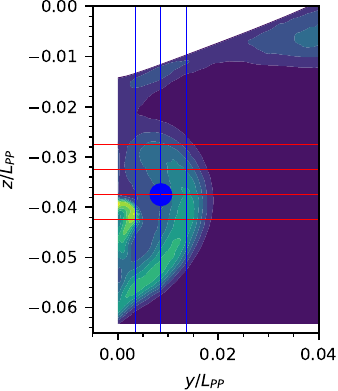}
    \label{fig:15e}}
    \subfloat[\textbf{2P} at \textbf{S7}]
	{\includegraphics[width=.33\linewidth]{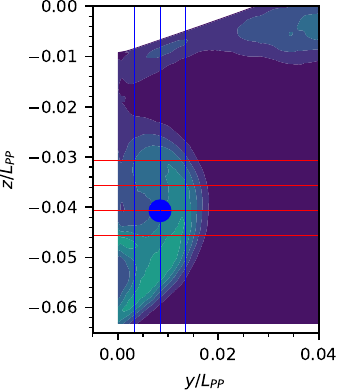}
    \label{fig:15f}}
    \caption{Total TKE of fine mesh for 1P and 2P at different cross sections.}
    \label{fig:15}
\end{figure}

Regarding the mean axial vorticity the differences in distribution are not striking but perceptible. The lower part of the hook-shape is wider for 2P and in the upper part it is slightly more intense than 1P. Further away from the hull (both to the side and top), for 2P there is less vorticity. The pictures (fig.~\ref{fig:15}) portray the distribution of TKE, with the point of reference being the vortex core of the 1P fine mesh and the grid lines drawn with a distance of $y, z / L_{PP} = 0.005$ each. The following observations are made:

\begin{itemize}
	\item The TKE field with free surface (2P) is closer to the hull.
	\item The 1P cases show a higher amount of TKE in the lower part of the hook shape. While 2P has less TKE in the lower part, there is a little more in the middle part and slightly more in the top part.
	\item The 2P formation sprawls further upwards.
\end{itemize}

\FloatBarrier


\section{Discussion}

\subsection{Validation, first and interim results}

It is demonstrated, that SLH works with free surface and both the medium and fine mesh are sufficiently resolved at the interface to capture the wave elevation in accordance with the experiments. It can be shown that far more than $80\%$ of TKE were directly resolved in the area of interest.


\subsubsection{TKE}
\label{sec:dis_1_TKE}

Regarding the very first TKE results, the substantial deviations from EFD and among each other cast doubt.
The very first idea is a drawback by the hybrid method, the grid-induced separation. On ambiguous grids it may happen that LES is activated already within the boundary layer due to sufficiently small cells, while the cell size at the same time is not adequate to resolve the full LES content. The subsequent reduction of the eddy viscosity and thus the Reynolds stresses is called modelled stress depletion (MSD). As a consequence premature separation is caused \citep{SPA06}. To get an idea if GIS is the problem, two parameters are investigated: circulation and resolved TKE.\\
The author would like to briefly anticipate that this investigation did not produce the desired results, which is why thanks to fruitful discussions with colleagues at the "IWSH 2023" conference, additional simulations were carried out. If it was GIS, the changes should show a correspondingly different result. The following was tried: First, the SLH criterion is doubled, delaying the switch to LES. The advantage is a larger URANS area close to the hull, while on the other side less LES inside the vortex core is expected. Second, a simulation with DDES will be conducted to have a comparison with an established method. Third, a single-phase mesh is created with a cell size between \enquote{medium} and \enquote{fine}, to determine a tendency -- if existing -- of the results regarding grid size. Fourth, an exemplary single-phase mesh with $y^+ < 0.1$ is created to completely avoid the usage of wall functions. This way it is possible to obtain reliable results especially for curved surfaces for all directions, as the wall function based on Prandtl only gives the tangential component. Last, the runtime is nearly doubled, as experience has shown that TKE needs longer to converge than other flow parameters. The results are discussed in section \ref{sec:dis_1_testSim}.

\subsubsection{Circulation}

Having a look at the vorticity fields and the circle positions of all meshes, it is revealed that
\begin{itemize}
	\item the position of the center point vastly influences the parts which will be considered at all for the final result,
	\item the results are massively distorted by positive parts which actually belong to the separation area.
\end{itemize}

While this is a well-known drawback using vorticity, the actual structure identified by the $\lambda_2$-criterion is smaller than when positive axial vorticity is used. Still, at S2 and S4, parts of the separation layer would be included when increasing the radius. Thus, the $\lambda_2$-criterion is not an option to use to exclude parts during the calculation of circulation.\\
The chosen radius is sufficient to capture the main vortex structure. The different starting positions due to the vortex centers deeply impact the results to the point where a comparison/ interpretation of the graphs cannot be done without considering additional information like the vorticity field. The axial vorticity behaves as expected: increasing with smaller grid size. Thus, despite some imponderabilities, neither the mesh set-up (regarding GIS) nor the method (SLH) seem to be the reason for the decrease in TKE for the fine meshes.

\subsubsection{Test simulations}
\label{sec:dis_1_testSim}

A doubled criterion means that the switch to LES is delayed; the URANS area at the hull is now thicker and less cells are calculated with pure LES. While the level of resolved TKE is still acceptable, this approach is not very sustainable, as it would be \enquote{try-and-error} for each simulation set-up. Effectively, doubling the criterion is like coarsening the mesh, thus the result does not really answer why the fine mesh produces this outcome.\\
Especially the fine mesh with $y^+ \leq 0.1$ was expected to show a different behaviour, as it theoretically could be run on pure LES and gets rid of possible disadvantages of wall functions.
Also, a closer look was given to cross sections in front of S2, to find out if e.g. a (premature) separation occurs. All meshes showed similar pictures for different parameters, of course with small differences, but similar in character (intensity, distribution). The differences seem to happen very sudden and could not be traced back to e.g. a certain position or change in geometry.\\
To sum up: while the results of the fine mesh(es) point to a typical problem of hybrid methods, namely GIS, the check of other related parameters and test simulations cannot confirm this presumption.\\
Regarding the actual research subject, the influence of the free surface on TKE, the results make it difficult to draw a conclusion. Especially for the medium mesh a tendency is noticeable, with 2P-TKE being larger at S4 and S7. For S2 the relation switches inside the vortex core. As the water level between S2 and S7 changes, one could search for an explanation there, regarding the tendency, but as the fine mesh shows either a different tendency or none at all, no conclusion will be drawn yet.\\
To get an additional impression of the reliability of the results, more simulations are carried out, this time with URANS. While it is clear that this will not yield useful results regarding TKE, the goal is to show the influence of the mesh on the parameters. Eventually it will help to evaluate if the influence of the free surface is more significant than the difference due to grid size.

\subsubsection{URANS}

The large differences between mesh sizes are not really explainable, as the vector and scalar fields are highly alike, and generally not expected for URANS. Already during the generation of the first diagrams, it was found that the position of the vortex cores varied between the simulations. That is why in the next step a closer examination of the position of the vortex cores is done, as well as a careful deliberation how reasonable the Tokyo'15 approach is.

\subsubsection{Vortex Cores}
\label{sec:dis_1_vortCore}

As no tomographic measurement data was available, it was chosen to use the maximum axial vorticity instead \citep[p. 187]{HIN21}. While this criterion works well for a developed flow which mainly exists parallel to the longitudinal axis, it is less reliable for slanted vortex structures, or if more than one (main) vortex exists. The vast majority of submissions did not utilize scale-resolving methods, resulting in one sole main vortex, and consequently one unambiguous center point. Researchers who used hybrid LES/ URANS could show the existence of small vortical structures which resulted in equivocal time-averaged patterns \citep{VIS16}.\\
While for the medium mesh two or three main vortices can be identified, clearly separated from each other, the vorticity field on the fine mesh looks almost DNS-like. The local maxima are substantially more numerous, and the difference between the values is tiny. While for the other meshes the maxima are in a similar range, the position of $\Omega_{x, max}$ of the fine mesh differs noticeably. However, for the sake of reliability and reproducibility, it was decided to stick to the procedure for the presented results. Comparing the position of the maximum axial vorticity with regard to the TKE field, it actually matches the relations between those two of the other meshes. Hence, the reason for the position shift and consequent results has to be something else and will be discussed later.

\begin{figure}[!htp]
	\centering
	\subfloat[Position of vortex cores at \textbf{S2}.]
	{\includegraphics[scale=1]{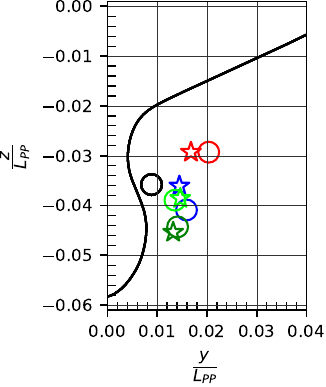}
    \label{fig:10a}}
    \subfloat[Position of vortex cores at \textbf{S4}.]
	{\includegraphics[scale=1]{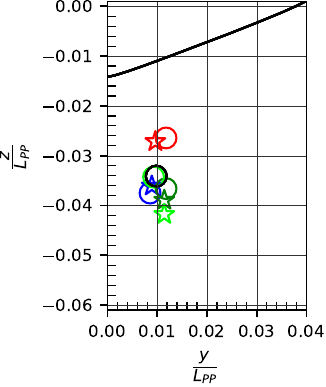}
    \label{fig:10b}}
    \subfloat[Position of vortex cores at \textbf{S7}.]
	{\includegraphics[scale=1]{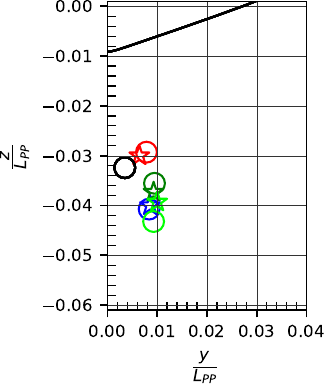}
    \label{fig:10c}}\\
    \subfloat
	{\includegraphics[scale=.8]{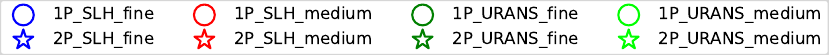}}
    \caption{Position of vortex cores at S2, S4 and S7.}
    \label{fig:10}
\end{figure}


First, let us have a look at the experimental results, depicted as black circle: especially at cross section S2 the position of the EFD vortex core is questionable. It is utmost close to the hull and from experience rather implausible. Several indications point to the position being wrongly detected:

\begin{itemize}
	\item Results from researchers submitted for Tokyo'15 do not coincide with the EFD location of the vortex core at S2, while being similar further away. Also the principal tendency of the curves is entirely opposing the EFD results \citep[p. 121]{HIN21}.
	\item From the EFD diagrams showing $U_{xx}$ one can gather that pretty much at the location of the EFD vortex core at S2 there is a small closed iso-contour with a higher velocity, but it is mentioned, that so close to the wall the PIV measurements might be less reliable, particularly as immediately next to the contour no more measurements were possible \citep[p. 161, p. 192]{HIN21}. It is unclear how the core position was determined for S2, if NMRI solely relied on $\Omega_{x,max}$ or also considered other criteria.
	\item The data which serves as comparison for the TKE diagram has a huge spike on the left side which no other participant found, at least not remotely that intense. It indicates the boundary layer gets touched where the measurement becomes unreliable/ is stopped.
\end{itemize} 

Coming back to the calculated results, the vortex center positions are fairly scattered, especially of the two medium meshes (fig.~\ref{fig:10}). Due to the uncertainty of the EFD outcome and other remaining open questions, no conclusion can be drawn yet.


\subsubsection{Fixed Position}

Surprisingly, the convictions need to be revised, as the adjusted results suggest considerably different outcome. First, there is a clear indication the fine mesh is not wrong, just shifted a little. If the shift was adjusted, the fine mesh would meet EFD much better than the medium mesh. Comparing URANS with the SLH results, the difference between the medium and fine SLH meshes is still significant, implying that supposedly the medium mesh is flawed. After various reasons have already been ruled out (see \ref{sec:dis_1_testSim}), it is now likely that either the area of interest is not resolved finely enough, or the area in front of the main vortex is too coarsely meshed.

\subsubsection{Change of medium mesh}

\begin{wrapfigure}{l}{.5\textwidth}
	\includegraphics[width=.5\textwidth]{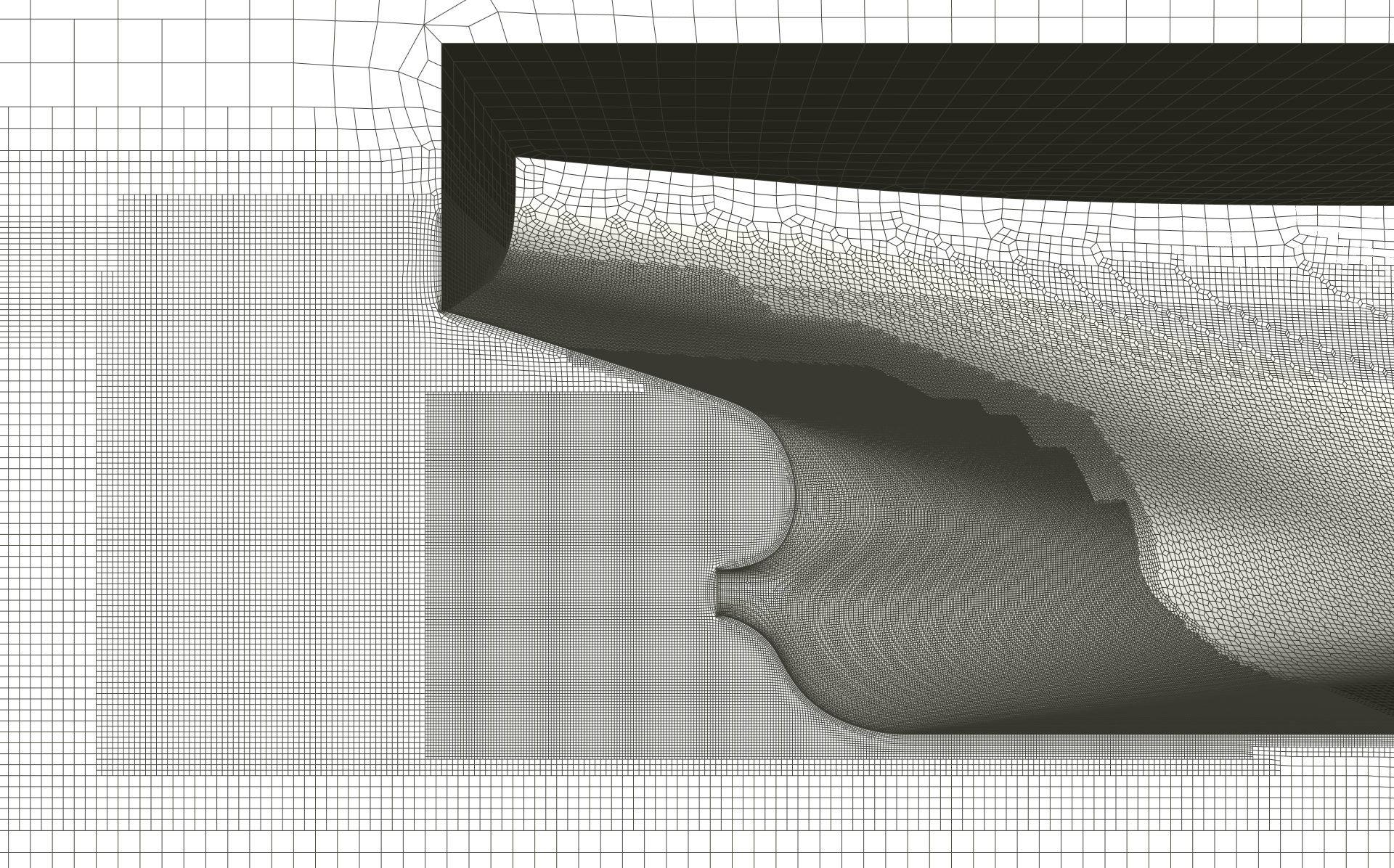}%
	\caption{Extension of finest refinement region on medium mesh.}
	\label{fig:12}
	\vspace{-1em}
\end{wrapfigure}

Following the precedent findings, two new meshes are generated:
\begin{enumerate}
	\item medium mesh with extra local refinement capturing the area where the main vortex evolves and exists, whereas the cell size of the local refinement corresponds to that of the fine mesh
	\item medium mesh where the refinement region of the smallest cell size is extended further forward (fig.~\ref{fig:12}).
\end{enumerate}

Due to the local refinement not showing any significant change compared with the original medium mesh, the results are not displayed. Apparently, the reason for the different results is not an insufficient resolution of the vortex structure.

\begin{figure}[!htp]
	\centering
	\subfloat[Position of vortex cores at \textbf{S2}.]
	{\includegraphics[scale=1]{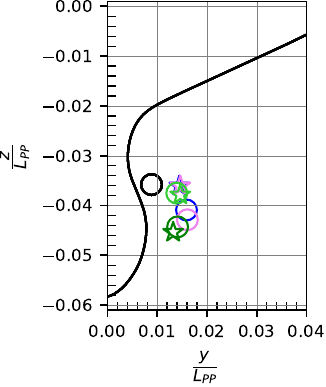}
    \label{fig:19a}}
    \subfloat[Position of vortex cores at \textbf{S4}.]
	{\includegraphics[scale=1]{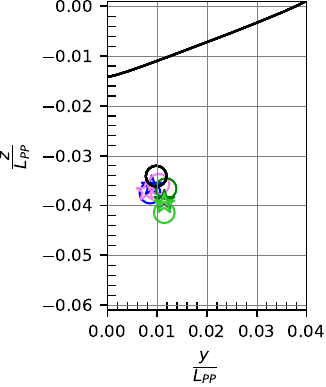}
    \label{fig:19b}}
    \subfloat[Position of vortex cores at \textbf{S7}.]
	{\includegraphics[scale=1]{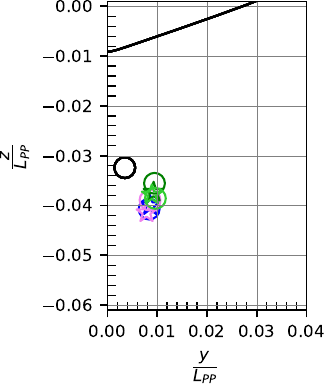}
    \label{fig:19c}}\\
    \subfloat
	{\includegraphics[scale=.8]{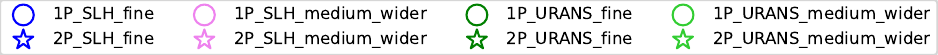}}
    \caption{Position of vortex cores at S2, S4 and S7 with new medium meshes.}
    \label{fig:19}
\end{figure}

On the other hand, extending the finest refinement region a bit to the ship front (not increasing the smallest cell size itself, thus keeping the characteristics of the medium mesh), the results are promising.\\
It seems, that although the meshes were compared thoughtfully beginning at cross sections where not yet the separation started and nothing visually pointed to that, some information was lost due to a too coarse mesh. To save cells and computation time/ resources, the extension is done goal-oriented and capturing around the double of the hull area than before, covering especially the part where the hull devolves into stern frames of high curvature. A test is carried out, if something changes when the second-finest refinement box (which covers the finest one and extends into all directions) is set to the smallest cell size (of the medium mesh), but the result does not change noticeably.\\
The new grids are now subjected to the same analyses as in the previous simulations. Exemplary the position of the vortex cores is shown again (fig. \ref{fig:19}). The graphics show a clear difference compared to figure \ref{fig:10}. Especially for the SLH simulations it is discernible that the new medium mesh is no longer a conspicuous outlier, but lies within the field of the other meshes.
It can also be noted that the positions now largely overlap with those of the fine mesh. Locations of both URANS and SLH simulations are now more clustered, with increasing congruity further away from the ship, which is attributed to the pronounced development of the bilge vortex.\\
To sum up the findings of this and the previous subsection, contrary to first assumptions, the fine mesh produces correct results. This also explains why measures to expose e.g. GIS did not yield anything.\\
Resulting from this, the final results and findings are done with the adjusted meshes.

\subsection{Final results}

With the new mesh and a single fixed position of the vortex core for all simulations, the results for TKE now are in much better agreement with EFD and between mesh sizes, especially inside the vortex core. Before one gets carried away interpreting a trend into something, the influence of the choice of the vortex center should be kept in mind. Neither Tokyo'15 methods nor using a fixed position allow for a reliable answer. That is why additionally the integral quantity of TKE for different radii (corresponding to the range of EFD measurements) is investigated.\\
The presentation of integral TKE depicts a quite clear picture. Having a closer look at the distribution of the TKE (see \ref{sec:D_distrTKE}), it is apparent that at S2 the positions of the vortex centers (see fig. \ref{fig:19}) of both the new medium and fine meshes are furthest apart from each other, compared with cross sections S4 and S7 where the coordinates are clustered close together. For the smaller radii this results in 2P being significantly greater than 1P, as due to fixing the position it is now placed closer to the inside of the TKE shape and thus higher values.\\
Nevertheless, it is important to note that for a bigger radius, which covers a large amount of the vortical structure, these differences become very small. That means, that overall the integral amount of TKE shows no difference between 1P and 2P meshes, and if so, it is smaller than possible errors due to different grid sizes.\\
Different findings are made when looking at the spatial distribution of TKE in the fluid. The shift closer to centerline of the 2P simulations is likely due to the oscillations of the transversal waves, which induce a clockwise motion, contrary to the main vortex structure which is spinning counter-clockwise. This leads to the whole formation being pressed against the hull, and also might be the reason why more content of TKE is in the upper half, compared with 1P.\\
It is important to mention that the procedures which was followed and the presented diagrams are constrained to the main vortex core and its direct neighbourhood.\\
Getting back to the literature presented at the beginning, both \textit{Maheo} and \textit{Kahraman et al.} investigated bodies which are very long compared to their width, and, contrary to a ship, have a large depth (plate and pier). Also, \textit{Maheo} and \textit{Babanin et al.} found the effects mainly close to the surface, as turbulence decays very fast with depth. These distinctions may explain that there is not a more striking effect in the results.


\section{Conclusions and future work}

A scale-resolving hybrid URANS/ LES method (SLH) was used to compare the turbulent kinetic energy in the wake of a full ship with and without free surface.
\begin{itemize}
	\item There are no indications for the JBC model at the given Froude number, that the overall amount of TKE inside and around the vortex center is influenced by the free surface.
	\item However, although the integral amount of TKE does not show significant differences, the spatial distribution varies noticeably. While this effect is negligible for bare-hull or resistance simulations, it is important for cases where e.g. structural (unsteady) loads are an essential part of the investigation. Considering a propeller or ESD, the free surface might substantially shift the position of pressure peaks.
	\item For the aforementioned cases, the use of scale-resolving methods and consideration of the free surface is recommended.
	\item Moreover, it is found to be difficult and not reliable to only use scalars in 2D plane to analyse 3D phenomena, wherefore it is recommended to use integral quantities instead/ additionally.
\end{itemize}

Further calculations are necessary to determine in how far these findings are valid for cases other than a full ship in calm deep water, like more slender ships or different stern geometries, shallow water, or higher velocities. As a next step, simulations will be set up to analyse the specific effect of the free surface on pressure peaks at defined points on a propeller or ESD.


%


\printcredits


\section*{Acknowledgements}

The author would like to thank Robinson Perić (TUHH) and Daniel Klembt (Corvus Works GmbH) for their comments on this manuscript and generously sharing their expertise.


\bibliographystyle{cas-model2-names}

\bibliography{dings_CLEAR}




\end{document}